\documentclass[pra,twocolumn,10pt,nofootinbib]{revtex4} 
\usepackage{amsmath,amssymb,graphicx}
\usepackage{color}

\usepackage{amsmath,amssymb}

\newcommand{\ket}[1]{|#1\rangle}
\newcommand{\bra}[1]{\langle#1|}

\definecolor{darkblue}{rgb}{0, 0, 0.8}
\usepackage[colorlinks=true, breaklinks=true, linkcolor=darkblue, citecolor=darkblue, urlcolor=darkblue]{hyperref}
\newcommand{\doilink}[2]{\href{http://dx.doi.org/#1}{#2}}

\usepackage{hyperref}

\begin{document}

\title[Experimental investigations of the dipolar interactions between single Rydberg atoms]{Experimental investigations of the dipolar interactions between a few individual Rydberg atoms}

\author{Antoine Browaeys, Daniel Barredo and Thierry Lahaye}

\address{Laboratoire Charles Fabry, Institut d'Optique, CNRS, Univ Paris Sud 11,
2~Avenue Augustin Fresnel, F-91127 Palaiseau Cedex, France}

\date{\today}

\begin{abstract}
This review summarizes experimental works performed over the last decade by several groups on the manipulation of a few individual interacting Rydberg atoms. These studies establish arrays of single Rydberg atoms as a promising platform for quantum state engineering, with potential applications to quantum metrology, quantum simulation and quantum information.
\end{abstract}

\maketitle

\section{Introduction}

Rydberg atoms~\cite{Gallagher1994} are highly excited atoms, where a valence electron has a large principal quantum number $n\gg 1$. They have exaggerated properties, and in particular they interact very strongly with each other via the dipole-dipole interaction. This is the basis for the \emph{Rydberg blockade}, i.e. the inhibition of the excitation of ground-state atoms to the Rydberg state by the presence of a nearby Rydberg atom. 

Early theoretical proposals~\cite{Jaksch2000,Lukin2001} suggesting to use the blockade to create entangled states of neutral atoms and quantum gates triggered a lot of experimental activity, over the last decade, to observe the blockade in ensembles of laser-cooled atoms~\cite{Saffman2010}. The field is now evolving along many directions, from quantum optics, with the promise to realize single-photon nonlinearities~\cite{Pritchard2013}, to many-body physics in large ensembles~\cite{Comparat2010}. This paper reviews recent experimental work on the Rydberg blockade and its application to the entanglement of two atoms as well as on the measurement of interactions between Rydberg atoms. We focus on small, well controlled systems of a few  individual atoms trapped in arrays of addressable optical tweezers~\cite{Schlosser2001}. We will only briefly mention recent works based on individual atoms held in optical lattices that use quantum gas microscopes~\cite{Schauss2012}.

This review is organized as follows. We first recall the motivation behind those studies, and in particular the principles of the quantum gates based on the blockade mechanism. Then, after a theoretical reminder about interactions between Rydberg atoms, we introduce the basic experimental techniques used to manipulate individual Rydberg atoms. We then review experiments that demonstrated the Rydberg blockade, quantum gates and entanglement of two atoms, and the direct measurements of the  interactions between Rydberg atoms in various regimes. Finally, we discuss the current efforts aiming at extending those studies to larger numbers of atoms.

\section{Motivation: individual Rydberg atoms for quantum state engineering}

\subsection{Review of single-particle Rydberg physics}

We first briefly recall some basic properties of Rydberg states and their scaling with the principal quantum number $n$ (see Table~\ref{tab:prop}). A comprehensive review of single-particle physics of Rydberg states can be found in~\cite{Gallagher1994}; short overviews are available in~\cite{Comparat2010,Loew2013}. As all the experiments using individual atoms performed to date use rubidium or cesium, our discussion is restricted to alkali atoms. 

Rydberg atoms are in states with a principal quantum number $n\gg 1$. This corresponds classically to a very large electron orbit, and the effect of the nucleus and remaining electrons (the ionic core) is essentially that of an elementary positive point charge: thus the properties of Rydberg atoms are very close to the ones of hydrogen. In particular, the energy of a state $\ket{n,l,j,m_j}$  is given by
\begin{equation}
E_{n,l,j}=\frac{-{\rm Ry}}{(n-\delta_{lj})^2}
\end{equation} 
where ${\rm Ry}\simeq 13.6\;{\rm eV}$ is the Rydberg constant, and the \emph{quantum defects} $\delta_{lj}$ are species-dependent corrections accounting for the effects of the finite size of the ionic core (for heavy alkali atoms, $\delta_{l\ge3}\approx 0$).

The typical size of the electronic wavefunction for a state $\ket{n,l,j,m_j}$ is on the order $n^2 a_0$, where $a_0$ is the Bohr radius. This size reaches hundreds of nanometers for the values of $n$  used in experiments (typically from $n\sim20$ to $100$), and is at the origin of the exaggerated properties of Rydberg states: the electric dipole matrix element between two neighboring states scales as $n^2$, while the energy spacing between adjacent Rydberg levels, which scales as $n^{-3}$, corresponds to millimeter-wave transitions. This gives the Rydberg atoms a long lifetime $\tau\sim n^3$, and a very strong sensitivity to electric fields: the polarizability scales as~$n^7$. This means that two nearby Rydberg atoms undergo very strong dipole-dipole interactions, that can reach tens of MHz for separations of several microns between the atoms.  

\begin{table}[b!]
\begin{center}
\begin{tabular}{ccc}
\hline
Property & $n-$scaling & \quad Value  (Rb $80S_{1/2}$) \\
\hline \hline
Binding energy $E_n$ & $n^{-2}$ & $-500$ GHz \\
Level spacing  $E_{n+1}-E_n$ & $n^{-3}$ & $13$ GHz \\
Size of  wavefunction $\langle{r}\rangle$ & $n^2$ & 500~nm \\
Lifetime $\tau$ & $n^3$ & 200~$\mu$s \\
Polarizability $\alpha$ & $n^7$ & $-1.8 \;{\rm GHz}/({\rm V/cm})^2$ \\
van der Waals coefficient $C_6$ & $n^{11}$ & 4~${\rm THz}\cdot\mu{\rm m}^6$ \\
\hline 
\end{tabular}
\end{center}
\caption{Properties of Rydberg states.}
\label{tab:prop}
\end{table}

Effects of Rydberg-Rydberg interactions were experimentally observed in 1981~\cite{Raimond1981}, at a time when Rydberg atoms were used as a test bed for the study of atom-light interactions~\cite{Haroche2013}. The interest in interacting Rydberg atoms got renewed at the end of the nineties, due to the novel possibilities offered by the availability of laser-cooled samples in which the atomic motion is negligible on relevant experimental timescales, thus realizing a `frozen Rydberg gas'~\cite{Anderson1998,Mourachko1998}. These pioneering studies motivated theoretical proposals~\cite{Jaksch2000,Lukin2001} suggesting to use the Rydberg blockade for quantum information processing~\cite{Saffman2005}. 

\subsection{Early proposals: Rydberg blockade and quantum gates}

\begin{figure}[t]
\centering
\includegraphics[width=6.9cm]{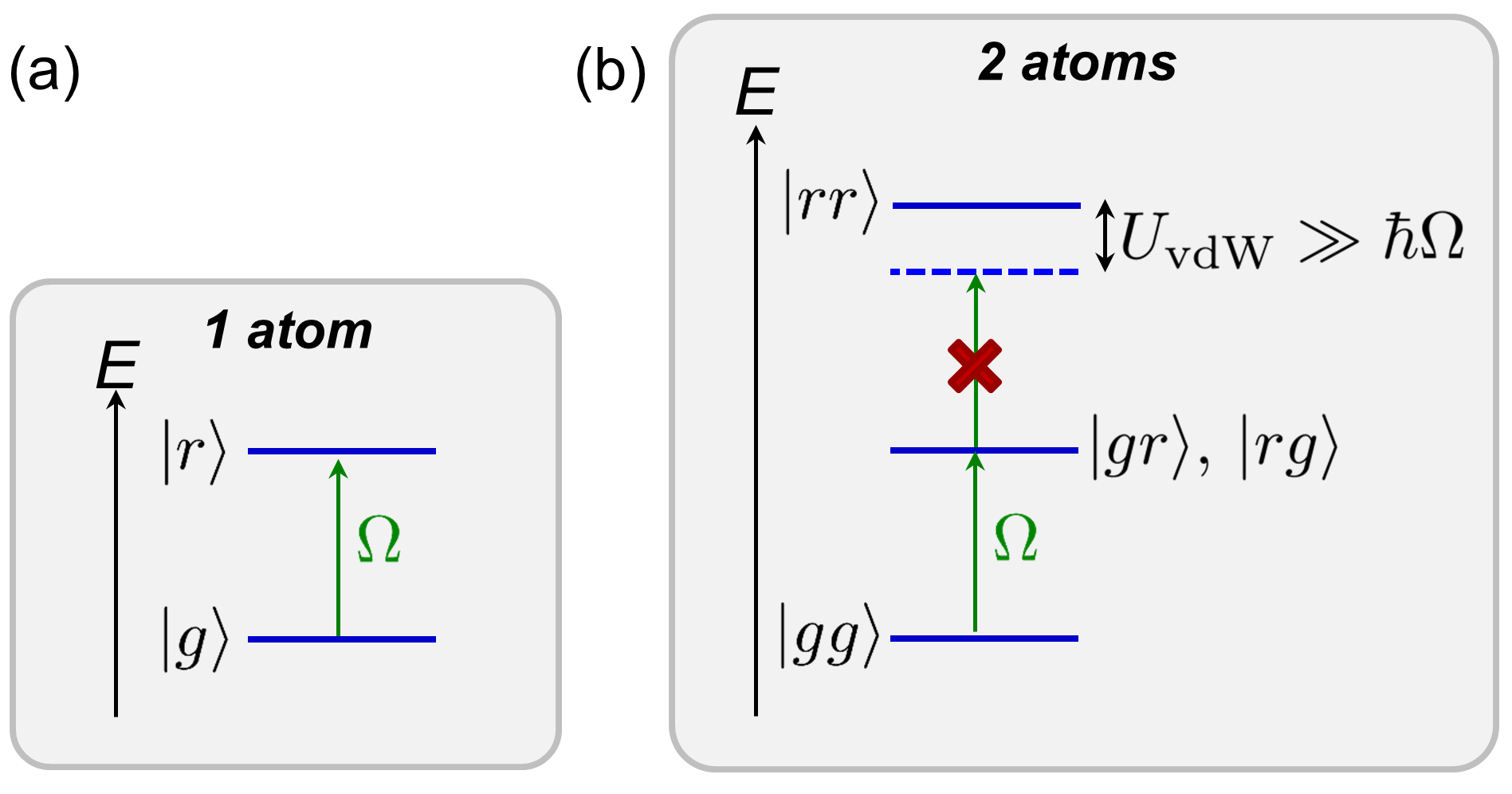}
\caption{Principle of the Rydberg blockade. (a) A resonant laser couples, with strength $\Omega$,  the Rydberg state $\ket{r}$ and the ground state $\ket{g}$ of an atom. (b) For two nearby atoms, interactions $U_{\rm vdW}$ shift the doubly excited state $\ket{rr}$, preventing the double excitation of the atom pair when $U_{\rm vdW}\gg \hbar \Omega$.}
\label{fig:blk_pple}
\end{figure}

The principle underlying the Rydberg blockade is shown in Figure~\ref{fig:blk_pple}. Consider the ground state $\ket{g}$ of an atom coupled to a Rydberg state $\ket{r}$ with a resonant laser with a Rabi frequency $\Omega$. In the case of two atoms, the collective ground state $\ket{gg}$ is still resonantly coupled to the states $\ket{gr}$ and $\ket{rg}$ containing a single Rydberg excitation. However, the doubly-excited state $\ket{rr}$ is shifted out of resonance by the strong van der Waals interaction $U_{\rm vdW}$ between the two atoms. In the limit $U_{\rm vdW}\gg \hbar\Omega$, i.e. for a small enough distance between the atoms, the double excitation is thus energetically forbidden: this is the Rydberg blockade\footnote{In the case of an incoherent excitation with a laser of linewidth $\gamma$, the blockade condition reads $U_{\rm vdW}\gg \hbar\gamma$.}. 

Introducing the two collective states $\ket{\psi_\pm}=\left(\ket{gr}\pm\ket{rg}\right)/\sqrt{2}$ we observe that the collective ground state $\ket{gg}$ is not coupled to $\ket{\psi_-}$, while its coupling to $\ket{\psi_+}$ is $\sqrt{2}\Omega$. Since $\ket{rr}$ is shifted out of resonance by the blockade condition, we end up with a two-level system comprising $\ket{gg}$ and $\ket{\psi_+}$, coupled by a \emph{collectively-enhanced} Rabi frequency $\Omega\sqrt{2}$. Starting from $\ket{gg}$ and applying the laser for a duration $\pi/(\Omega\sqrt{2})$  thus prepares the entangled state $\ket{\psi_+}$\footnote{Strictly speaking, if ${\boldsymbol r}_1$ and ${\boldsymbol r}_2$ denote the positions of atoms 1 and 2, the entangled state $\ket{\psi_+}$ reads $\left({\rm e}^{i {\boldsymbol k} \cdot {\boldsymbol r}_1}\ket{rg}+{\rm e}^{i {\boldsymbol k} \cdot {\boldsymbol r}_2}\ket{gr}\right)/\sqrt{2}$, where ${\boldsymbol k}$ is the wavevector of the laser field coupling $\ket{g}$ to $\ket{r}$. For simplicity, we will omit these phase factors in this review, unless in cases where they are important.}.

The above arguments extend to $N>2$ atoms if all pairwise interactions meet the blockade criterion, i.e. if all the atoms are contained within a `blockade sphere' of radius $R_{\rm b}=[C_6/(\hbar\Omega)]^{1/6}$ (this blockade radius can reach several microns for typical experimental parameters). One gets a collectively enhanced Rabi-oscillation at frequency $\Omega\sqrt{N}$ between the collective ground state $\ket{ggg\cdots}$ and the entangled W-state
\begin{equation}
\ket{W}=\frac{1}{\sqrt{N}}\left(\ket{rgg\cdots}+\ket{grg\cdots}+\cdots+\ket{gg\cdots r}\right),
\end{equation}
where a single Rydberg excitation is delocalized over all the atoms. 

\begin{figure}[t]
\centering
\includegraphics[width=8.5cm]{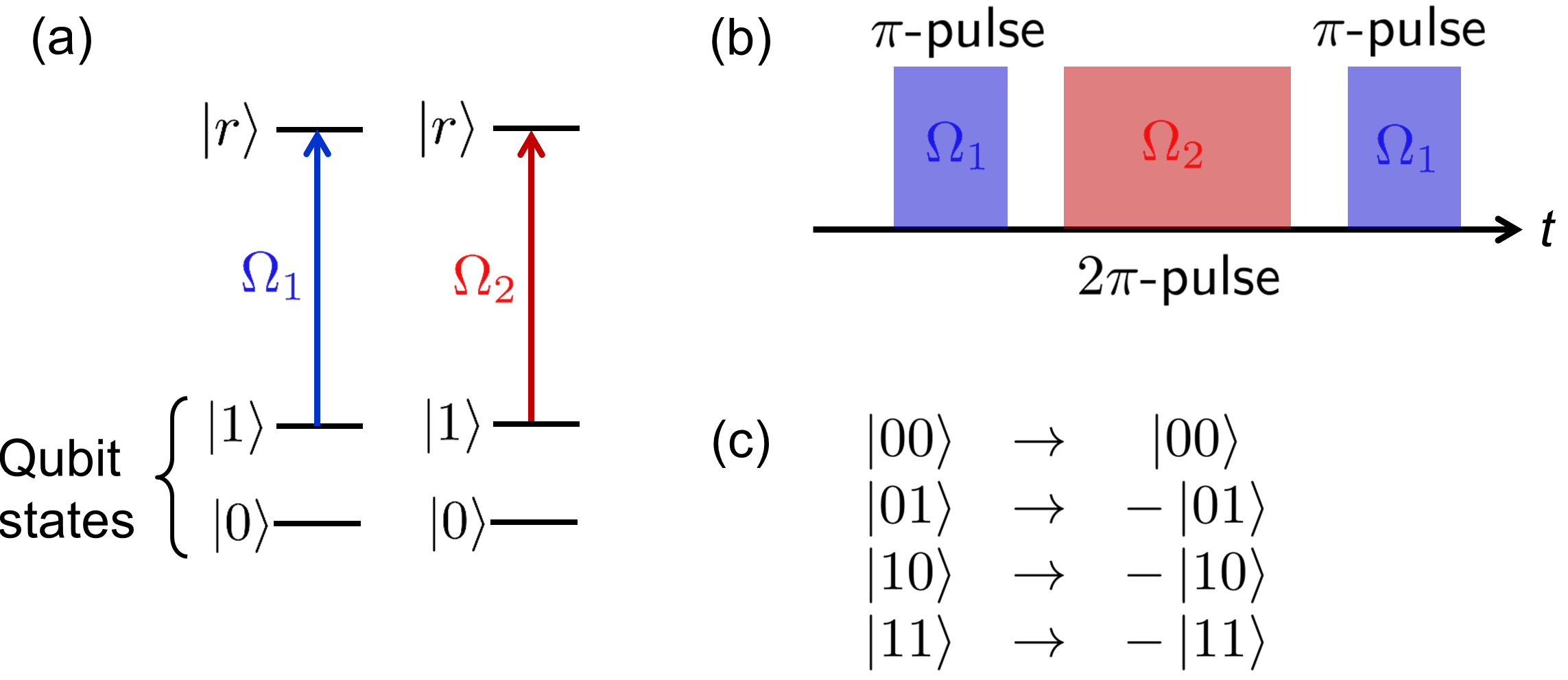}
\caption{Principle of a two-qubit quantum gate based on Rydberg blockade. (a) Involved levels and lasers. (b) Pulse sequence. (c) Truth table of the phase gate.}
\label{fig:gate}
\end{figure}

The Rydberg blockade was proposed in~\cite{Jaksch2000} as a means of implementing fast quantum gates with neutral atoms. The principle is shown in Fig.~\ref{fig:gate}. The qubits are encoded in two long-lived hyperfine levels $\ket{0}$ and $\ket{1}$ of the ground state of each atom, which can be separately addressed by lasers that couple state $\ket{1}$ to the Rydberg state $\ket{r}$ (Fig.~\ref{fig:gate}a). The two atoms are close enough so that Rydberg blockade prevents the excitation of $\ket{rr}$. When applying the pulse sequence shown in Fig.~\ref{fig:gate}b, if any of the qubit is initially prepared in $\ket{1}$, then the blockade makes one of the lasers off-resonant, one and only one of the atoms undergoes a $2\pi$ rotation, and the wavefunction of the system gets a minus sign at the end of the sequence. If both qubits are initially in $\ket{0}$, the laser pulses have no effect. This leads to the truth table shown in Fig.~\ref{fig:gate}c, which implements a controlled-phase gate (that can be turned into a CNOT gate using additional  single-qubit gates).  One appealing feature of the Rydberg gates lies in its short duration, set by the interaction energy of the two atoms: as it can be as large as 10 MHz, the gate can operate on a sub-microsecond time scale. This is in contrast with entangling operations using e.g. much weaker ground-state interaction~\cite{Mandel2003}, which  operate on a much longer time. Another strong other asset of this protocol is that it is largely insensitive to the exact value of the interatomic interaction.

Further theoretical studies proposed to use the Rydberg blockade in atomic ensembles~\cite{Lukin2001,Saffman2002} in order to generate non-classical states of light, or encode collective qubits. These early proposals were then followed by detailed theoretical analyses of the various sources of possible experimental imperfections~\cite{Saffman2005,Muller2014}, that showed promising prospects for the realization of high-fidelity gates. After the first demonstration of the blockade between two atoms (see Sec.\ref{Sec:blockade_exp}), new schemes where proposed for quantum gates~\cite{Saffman2009,Isenhower2011}, including a generalized CNOT gate where one atom controls the state of many others~\cite{Mueller2009}, or for the preparation of multi-partite entangled states~\cite{Moller2008}. 

\subsection{Quantum simulation}

Building a useful, general-purpose quantum computer is to date an extremely challenging task, due to the very large number of qubits and high-fidelity gates that are required~\cite{Nielsen2000}. A seemingly more realistic goal is to realize \emph{quantum simulators}~\cite{Feynman1982,Lloyd1996}, in particular analog ones, i.e. well-controlled quantum systems that can be used to realize physically, in the laboratory, a complex, many-body Hamiltonian of interest in other fields, e.g. in condensed-matter physics~\cite{Georgescu2014}. Interesting properties of the Hamiltonian, that are in practice impossible to obtain from theoretical or numerical studies, can then be directly measured in the simulated system. 

Rydberg atoms are attractive candidates for the realization of quantum simulators~\cite{Weimer2010}. In particular, as we shall see in the next section, the interactions between Rydberg atoms naturally implement analog simulations of various types of spin Hamiltonians, such as the Ising model or the XY model, where the spin states are encoded in different atomic levels. 

\section{Interaction between Rydberg atoms}\label{Sec:theo_interaction}

In this section, we briefly describe various regimes of interactions between two Rydberg atoms.  We restrict ourselves to a perturbative approach, and only outline the main 
features of the problem for the simple case of alkali atoms.  For details about actual numerical calculations, we refer for instance to~\cite{Reinhard2007}.

\subsection{Perturbation of pair states by the dipole-dipole interactions}

We consider two atoms, labeled $1$ and $2$, located at positions ${\boldsymbol R}_1$ and ${\boldsymbol R}_2$, and  we denote by ${\boldsymbol R}={\boldsymbol R}_2-{\boldsymbol R}_1$ their separation.  When $R\equiv|{\boldsymbol R}|$ is much larger than the size of the electronic wavefunction,  the interaction Hamiltonian is obtained by the multipolar expansion, and the dominant term is the dipole-dipole interaction 
\begin{equation}
V_{\rm ddi} = \frac{1}{4\pi\varepsilon_0}\frac{{\boldsymbol d}_1\cdot{\boldsymbol d}_2-3({\boldsymbol d}_1\cdot{\boldsymbol n})({\boldsymbol d}_1\cdot{\boldsymbol n})}{R^3},
\label{eq:vdd}
\end{equation}
with ${\boldsymbol n}={\boldsymbol R}/R$, and ${\boldsymbol d}_i$ the electric dipole operator of atom $i$.

\begin{figure}[t]
\centering
\includegraphics[width=8.5cm]{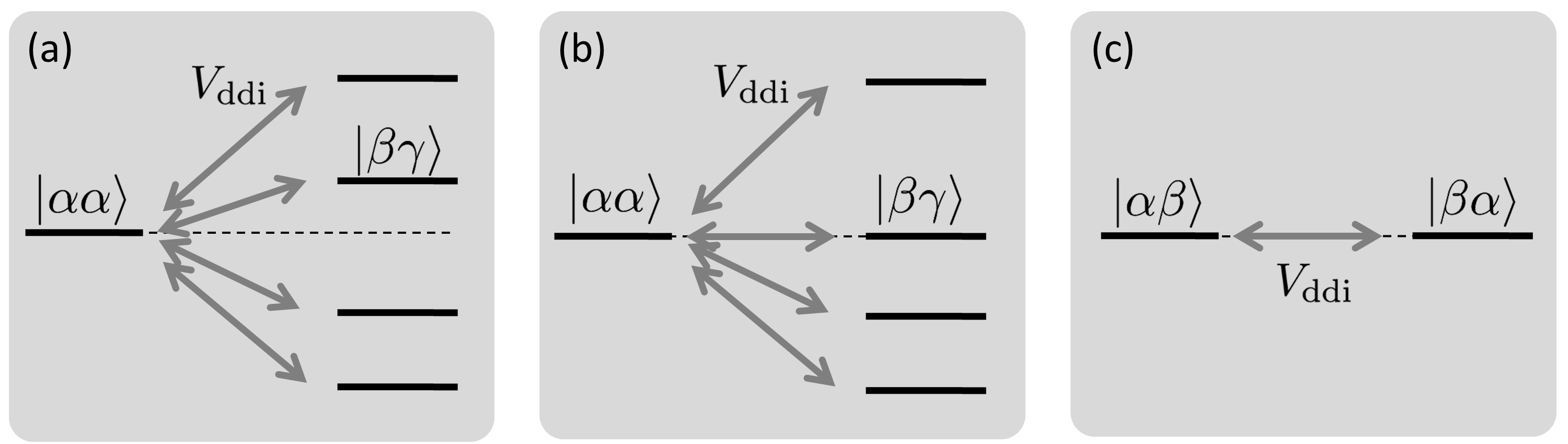}
\caption{Various types of interactions between two Rydberg atoms. (a) Van der Waals regime.  (b) F\"orster resonance.  (c) Resonant dipole-dipole interaction between two different Rydberg states $\ket{\alpha}$ and $\ket{\beta}$.}
\label{fig:inter}
\end{figure}

Let us denote by $\ket{\alpha},\ket{\beta},\ldots$ the eigenstates of a single atom, with  corresponding eigenenergies  $E_\alpha, E_\beta,\ldots$ ($\alpha$ summarizes the quantum numbers ${n,l,j,m_j}$). In the absence of interaction, the eigenstates of the two-atom system are the \emph{pair states} $\ket{\alpha\beta}\equiv\ket{\alpha}\otimes\ket{\beta}$ with energies $E_{\alpha\beta}=E_\alpha+E_\beta$. Our goal is to calculate the effect of the perturbation (\ref{eq:vdd}) on these pair states; depending on the situation, three regimes can be obtained (see Figure~\ref{fig:inter}).

\subsection{Van der Waals regime}

We first assume that the two atoms are prepared in the same state $\ket{\alpha}$. In general, the pair state $\ket{\alpha\alpha}$ is not degenerate with any other pair state (Figure~\ref{fig:inter}a), the typical splittings being several GHz. We thus use non-degenerate perturbation theory. To first order, there is no energy shift, as the average value of $V_{\rm ddi}$ in $\ket{\alpha\alpha}$ vanishes due to the fact that ${\boldsymbol d}_i$ is an odd-parity operator and that the atomic states $\ket{\alpha}$ have definite parity. The energy shift is thus given by second-order perturbation theory
\begin{equation}
\Delta E_{\alpha\alpha} = \sum_{\beta,\gamma\ldots}\frac{|\langle\alpha\alpha|V_{\rm ddi}|\beta\gamma\rangle|^2}{E_{\alpha\alpha}-E_{\beta\gamma}},
\label{eq:vdw}
\end{equation}
where the sum extends to all states that are dipole-coupled to $\ket{\alpha}$. Being second-order in $V_{\rm ddi}$, the shift scales as $1/R^{6}$ and is simply the van der Waals interaction. As the numerator in~(\ref{eq:vdw}) is proportional to a dipole moment to the fourth power, it scales as $n^8$; the denominator, being a difference in energy between adjacent pair states, scales as $1/n^3$. The $C_6$ coefficient thus increases dramatically with $n$, as $n^{11}$.

For a system of $N>2$ atoms, the effects of van der Waals interactions are pairwise additive (except in exceptional cases, in particular when one considers the van der Waals interaction between e.g. different states of the same parity~\cite{Cano2012}). Therefore, the interaction Hamiltonian for $N$ atoms reads
\begin{equation}
H_{\rm vdW}=\sum_{i<j}\frac{C_6}{R_{ij}^6}n_i n_j
\label{eq:vdw:natoms}
\end{equation}
where $n_i=|r\rangle\langle r|_i$ is the projector on the Rydberg state of interest of atom number $i$. If one introduces pseudo-spin $1/2$ states $\ket{\!\!\downarrow}=\ket{g}$, where $\ket{g}$ is the atomic ground state, and $\ket{\!\uparrow}=\ket{r}$, along with the corresponding spin operators $\sigma_{x,y,z}$, one can write $n_i=(1+\sigma_z^i)/2$.
When one adds a coherent laser driving on the transition $\ket{g}\leftrightarrow\ket{r}$ with a Rabi frequency $\Omega$ and a detuning $\delta$, the total Hamiltonian (in the rotating frame of the laser) is:
\begin{equation}
H_{\rm Ising}=\frac{\hbar\Omega}{2}\sum_i\sigma_x^i+\sum_i (\hbar\delta+B_i)\sigma_z^i+\sum_{i<j}\frac{C_6}{R_{ij}^6}\sigma_z^i\sigma_z^j,
\label{eq:ising}
\end{equation}
with $B_i=\sum_jC_6/R_{ij}^6$. 
In the language of spin Hamiltonians, (\ref{eq:ising}) describes an Ising 
quantum magnet with a transverse field $\propto \Omega$, a longitudinal field $\propto \hbar\delta+B_i$, 
and a spin-spin coupling decaying as $1/R^6$ with the distance $R$ between the spins.

For Rydberg states with an orbital angular momentum $L$, each atom has $2J+1$ degenerate (or almost degenerate in the presence of a moderate magnetic field) Zeeman sublevels (here, $J=L\pm1/2$ is the total angular momentum). This means that instead of having to consider a single isolated pair state $\ket{\alpha\alpha}$, one has to deal with a manifold of $(2J+1)^2$ states. They are not directly coupled with each other by (\ref{eq:vdd}), but second-order perturbation theory gives an effective Hamiltonian 
that acts within the manifold, with a global $1/R^6$ scaling and couplings that depend on the angle $\theta$ between the quantization axis and the internuclear axis. In the blockade regime, it is possible to define an effective van der Waals shift, given by a suitably weighted average of the eigenvalues of the effective Hamiltonian~\cite{Walker2008}. This allows one to keep a simple (but approximate) two-level description of each atom, keeping the size of the Hilbert space equal to $2^N$ for $N$ atoms~\cite{Vermersch2015a}. The validity of such approximations will depend on the exact experimental settings.

\subsection{F\"orster resonance: tuning the interaction with an electric field}

For some values of $n$, the pair state $\ket{\alpha\alpha}$ can be  degenerate or quasi-degenerate with another pair state $\ket{\beta\gamma}$ with which it is coupled by the dipole-dipole interaction (Figure~\ref{fig:inter}b). In this case, one neglects the other, non-resonant couplings, keeping two coupled degenerate states. Then the eigenstates in the presence of $V_{\rm ddi}$ are $\ket{\pm}=\left(\ket{\alpha\alpha}\pm\ket{\beta\gamma}\right)/\sqrt{2}$, and the eigenergies $E_\pm=\pm C_3/R^3$, where $C_3=R^3\bra{\beta\gamma}V_{\rm ddi}\ket{\alpha\alpha}$. The interaction is now resonant and scales as $1/R^3$~\cite{Gallagher1994}. Such resonances have been called F\"orster resonances~\cite{Saffman2005b,Beterov2015}, due to the analogy with the F\"orster resonance energy transfer~\cite{Forster1948,Clegg2006} at work in photochemistry.

Such degeneracies of pair states are in general only approximate, with a difference in energy  $\Delta=E_{\alpha\alpha}-E_{\beta\gamma}$ (called the \emph{F\"orster defect}) between the two quasi-degenerate pair states of a few or a few tens of MHz. However, $\ket{\alpha}$, $\ket{\beta}$ and $\ket{\gamma}$ have in general different polarizabilities, making it possible, by applying moderate electric fields, to Stark-tune the relative positions of $\ket{\alpha\alpha}$ and $\ket{\beta\gamma}$ in order to get to exact resonance. Experimentally, this allows one to switch, on fast timescales and almost at will, between (strong) resonant and (weak) non-resonant (van der Waals) interactions between the atoms. 

Due to the Zeeman substructure of the involved Rydberg levels, there are in general several resonances between different channels corresponding to the various possible combinations of the $m_j$ values. They occur at slightly different values of the electric field, and that have a different angular dependence due to the anisotropy of the dipolar interaction~\cite{Nipper2012a,Nipper2012b,Ravets2015}. 

For a fixed, non-zero F\"orster defect, one observes a transition between the F\"orster regime at short distances, and the van der Waals regime at large distances. The crossover between the two regimes occurs at a distance $R_c\sim (C_3/\Delta)^{1/3}$. Away from quasi-degeneracies, $R_c$ scales as $n^{7/3}$ with the principal quantum number $n$.

\subsection{Resonant dipole-dipole interactions: ``spin-exchange'' Hamiltonian}\label{Sec:theo_reson_dipdip}

Another possibility to observe resonant dipole-dipole interactions is to use two distinct, dipole-coupled Rydberg states, by preparing the pair in $\ket{\alpha\beta}$, where for instance $\alpha$ is a $nS$ Rydberg state, and $\beta$ a $n'P$ state (with $n\simeq n'$). The pair state $\ket{\alpha\beta}$ being degenerate with $\ket{\beta\alpha}$, and the dipole-dipole Hamiltonian (\ref{eq:vdd}) coupling these two states, $V_{\rm ddi}$ reduces to (in the basis $\{\ket{\alpha\beta},\ket{\beta\alpha}\}$):
\begin{equation}
V_{\rm ddi}=\frac{C_3}{R^3}\left(\ket{\alpha\beta}\bra{\beta\alpha}+\ket{\beta\alpha}\bra{\alpha\beta}\right),
\end{equation}
where the coefficient $C_3$ is the product of two matrix elements of the dipole operator between $\ket{\alpha}$ and $\ket{\beta}$ and therefore  scales as $n^4$.

From the point of view of quantum simulation, if one encodes pseudo-spin states $\ket{\!\!\!\uparrow},\ket{\!\!\!\downarrow}$ in $\ket{\alpha},\ket{\beta}$, the resonant dipole-dipole interaction directly implements the XY spin Hamiltonian:
\begin{equation}
H_{XY}=\sum_{i<j}\frac{C_3}{R_{ij}^3}\, \left(\sigma_+^{i}\sigma_-^{j} + \sigma_-^{i}\sigma_+^{j} \right),
\label{eq:XY_hamiltonian}
\end{equation}
with spin couplings decaying as $1/R^3$. Here, $\sigma_\pm=\sigma_x\pm i \sigma_y$. Such long-ranged spin Hamiltonians have been predicted  to display anomalous properties as compared to their short-range counterparts~\cite{Hauke2010,Peter2012}, making them attractive from the point of view of quantum simulation, and have been the subject of experimental studies using ultracold polar molecules pinned in optical lattices~\cite{Yan2013} or dipolar Bose--Einstein condensates~\cite{Paz2013}.

\begin{figure}[t]
\centering
\includegraphics[width=7cm]{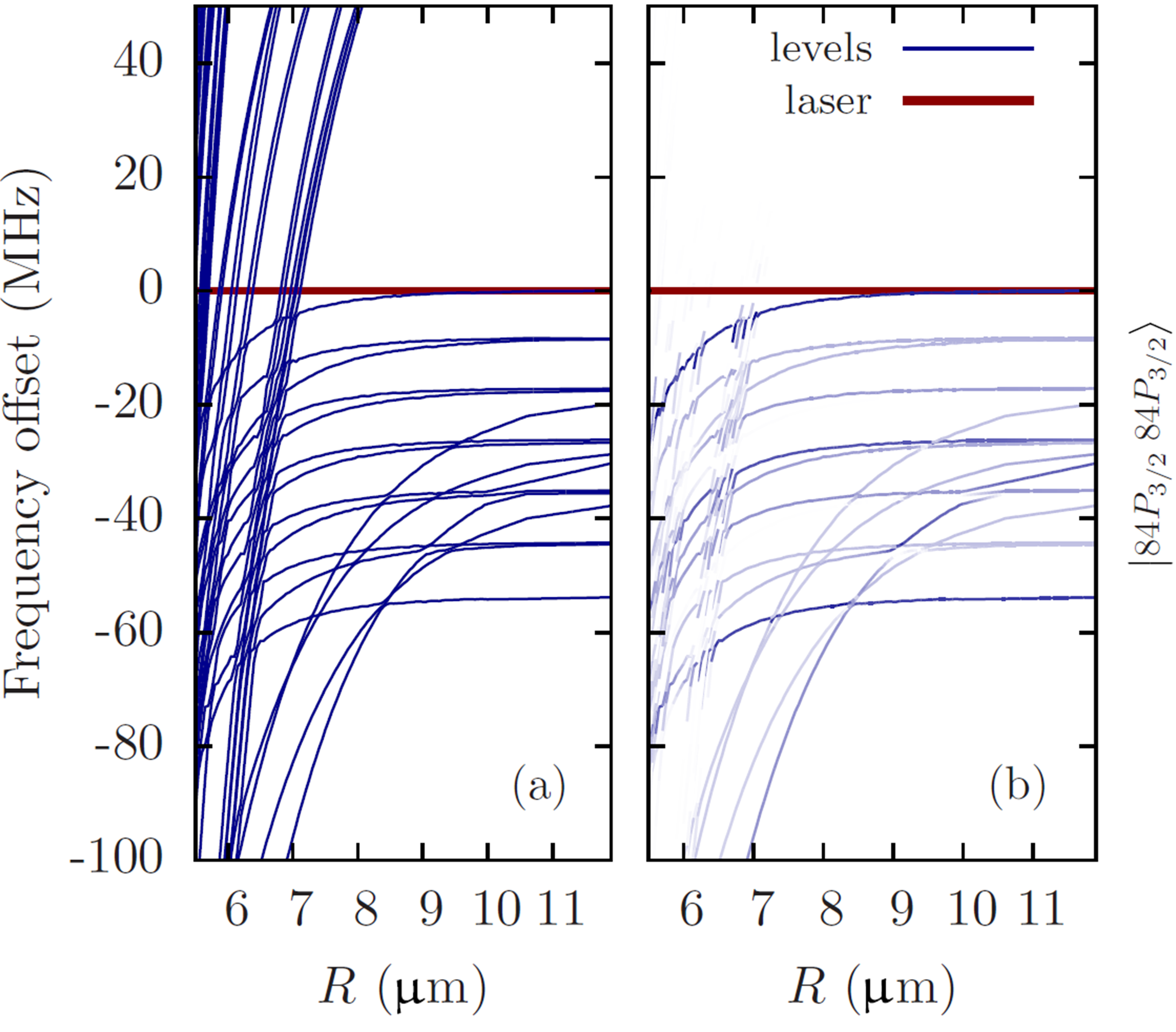}
\caption{A part of the spectrum of a system of two $^{133}{\rm Cs}$ atoms separated by a distance $R$ in the presence of the dipole-dipole interaction (and of electric and magnetic fields) obtained by numerical diagonalization.  (a) Full plot of the spectrum. (b) Same as (a) but with the darkness of the lines weighted by the oscillator strength to the ground state. Figure taken from~\cite{Hankin2014}.}
\label{fig:full_diag}
\end{figure}

\subsection{Beyond perturbation theory: numerical diagonalization of the Hamiltonian}

The discussions above give simple expressions for the effects of interactions on a pair of atoms. However, for accurate comparison with experiments, it is necessary to resort to a full numerical calculation of the energy spectrum of the pair, as the large number of closely-spaced Rydberg states for large $n$ leads to deviations from the simple van der Waals interaction even for shifts as small as a few tens of MHz. For this purpose, one needs to evaluate numerically the (radial) dipole matrix elements between different Rydberg wavefunctions, and thus, the wavefunctions themselves. This can be accomplished by solving the radial Schr\"odinger equation using the Numerov method~\cite{Zimmerman1979}. The (truncated) Hamiltonian comprising the single-atom part and the dipole-dipole interaction (\ref{eq:vdd}) is then diagonalized numerically (typically a few hundreds or a few thousands states are retained). Figure~\ref{fig:full_diag}a shows a typical result of such a calculation, showing that at distances of a few micrometers, many molecular states, with an energy varying very rapidly with the distance $R$, cross the line $\Delta E=0$ corresponding to non-interacting atoms. This might give the impression that blockade breaks down at short distances. However these states actually are very weakly laser-coupled to the ground state (see Fig.~\ref{fig:full_diag}b), which preserves the quality of the blockade (see also~\cite{Derevianko2015}).

\section{Experimental considerations: trapping and Rydberg excitation of individual atoms}

In this section we describe the main experimental tools used in recent experiments where arrays of single Rydberg atoms are exploited for quantum simulation and quantum information processing applications. Most experiments so far were performed by the University of Wisconsin (USA) group, the Sandia National Laboratory group (Albuquerque, USA), and the Institut d'Optique group (Palaiseau, France) using similar methods. We summarize here these experimental techniques for the preparation, detection, and manipulation of individual Rydberg atoms.

\subsection{Trapping individual atoms in ``optical tweezers''}

Neutral atoms can be confined in space by the conservative potential of a far-off resonance optical dipole trap~\cite{Grimm1998}. An optical dipole trap is formed by  focusing a laser beam tuned far away from the atomic resonance frequency. Red-detuned light induces an electric dipole moment in the atoms and exerts a force towards regions of maximal intensity. This creates effective potentials with typical depths $U /k_B \approx 0.1-1\, \rm{mK}$. To load the atoms in the traps a standard method is to pre-cool the atoms in a Magneto Optical Trap (MOT). Loading is achieved by spatially overlapping the dipole trap with the atomic reservoir created by the MOT, leading to mesoscopic ensembles of atoms with temperature $\sim 10 \,\mu \rm{K}$. 

Single atom trapping needs further requirements, and different approaches have been followed. One possibility is to prepare an exactly known number of atoms (1-30) in the MOT by setting its parameters to the limits (e.g. using a large magnetic field gradient)~\cite{Hu1994}, before transferring the atoms to the dipole trap~\cite{Kuhr2001}. This technique has been successfully applied to the generation of arrays of single atoms in one dimensional optical lattices~\cite{Meschede2006}.

\begin{figure}[t]
\centering
\includegraphics[width=7.5cm]{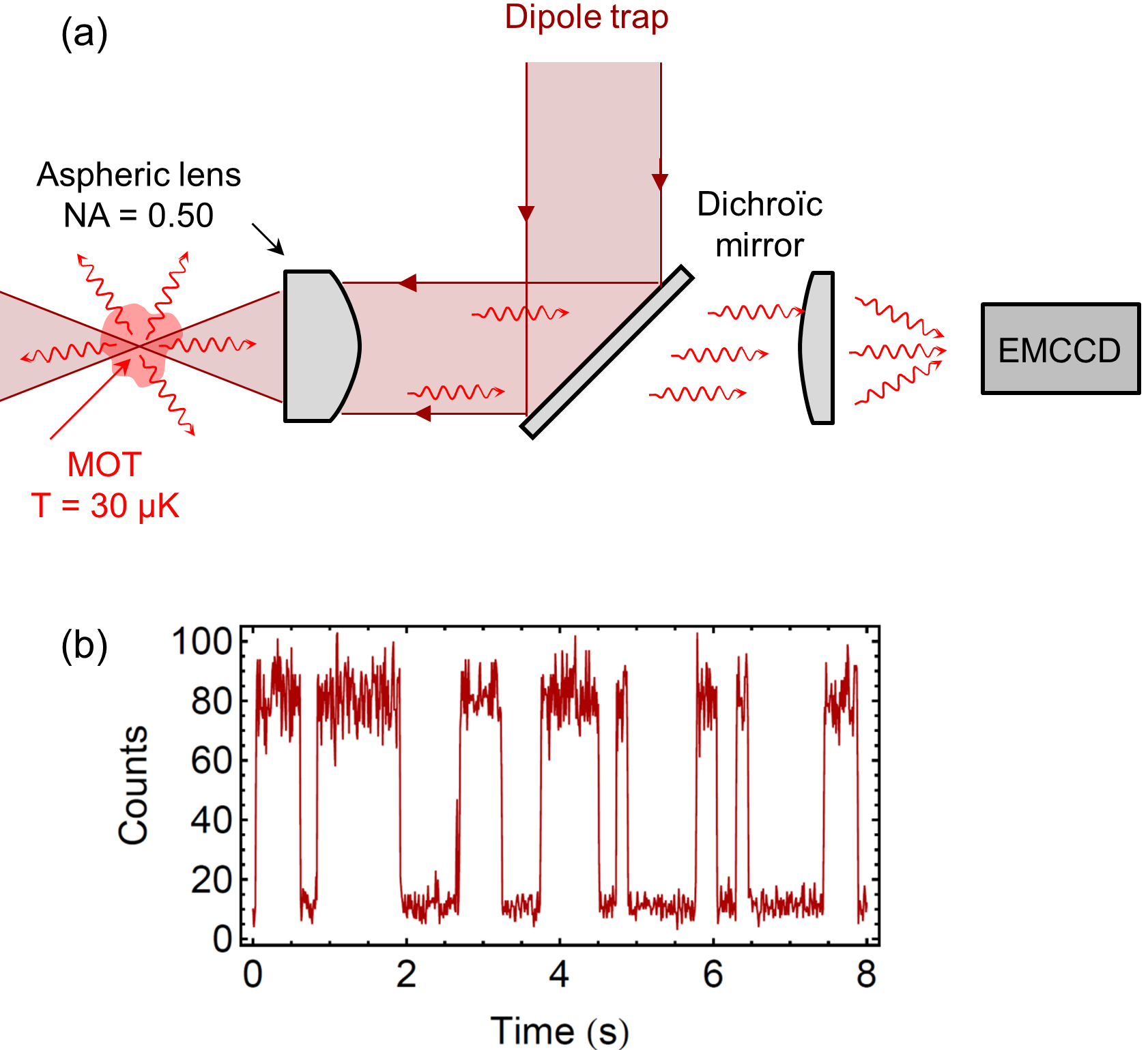}
\caption{Loading and imaging single atoms in a dipole trap. (a) Atoms initially trapped in a MOT are loaded in a dipole trap formed by focusing a red-detuned laser beam with a high numerical aperture aspheric lens (${\rm NA} = 0.5$) under vacuum~\cite{Sortais2007}. The fluorescence of the atoms is separated from the dipole trap light with a dichroic mirror and imaged on an EMCCD camera. (b) Single-atom fluorescence signal with two fluorescence levels, corresponding to one or zero atoms in the trap.}
\label{fig:single_atoms}
\end{figure}

A second possibility is to use a high numerical aperture optical system, such as custom-made objectives~\cite{Alt2002} or aspherical lens~\cite{Sortais2007}, to reduce the volume of the dipole trap to a size of $\sim 1 \,\mu\rm{m}^3$. This configuration of a tightly focused dipole trap is named ``optical tweezers".
In such a small trap, the dynamics of the atoms is governed by fast inelastic light-assisted collisions (with rates of $\sim 10^4 \mu\rm{m}^3 s^{-1}$) induced by the near-resonant MOT light~\cite{Fuhrmanek2012}, and is dominated by two-body losses~\cite{Schlosser2001,Schlosser2002}. As a consequence, there exists a regime of densities of the cold-atom cloud where the loading is sub-Poissonian and at most one atom is trapped at a time. In this regime a first atom of the cloud enters the tweezers and is slowed down thanks to the cooling lasers. When a second atom enters the tweezers, a two-body inelastic collision catalyzed by the light results in the rapid loss of the two atoms.

The configuration using a tight dipole trap presents the advantage of being easily combined with an imaging system with micrometer resolution, as represented in Fig~\ref{fig:single_atoms}a. In this way, 
a real-time imaging system can be used to record the fluorescence of the atoms when they are illuminated with near-resonant laser light (Fig.~\ref{fig:single_atoms}b): the fluorescence signal toggles at random between periods of low values corresponding to an empty trap, and periods of high value reflecting the presence of an atom. It is thus possible to determine exactly when an atom has entered the trap and use this information to trigger single-atom experiments with typically $<$ 1 s duty cycle. Table~\ref{tab:single_atom_properties} gives typical parameters for an individual atom trapped in an optical tweezers.  

\begin{table}[t]
\begin{center}
	\begin{tabular}{lr}
		\hline
		Quantity & Typical value \\
		\hline
		\hline
		Trap wavelength   & 852~nm \\
		Trap power   & 4~mW \\
		Trap beam waist (intensity, $1/e^2$)   & $ 1.1 \, \mu \rm{m}$ \\
		Trap depth  $U /k_B$  & 1~mK \\
		Longitudinal trap frequency $\omega_l\,$ & $ 2\pi\,\times$ 15~kHz \\
		Radial trap frequency  $\omega_r$ & $2\pi\,\times$ 90~kHz \\
		MOT temperature  & $100 \,\mu \rm{K} $  \\
		Single-atom temperature  & $30 \,\mu \rm{K} $  \\
		\hline 
	\end{tabular}
\end{center}
	\caption{Representative values for single-atom trapping in the experiments at Institut d'Optique (Palaiseau) using $^{87}$Rb.}
	\label{tab:single_atom_properties}
\end{table}

This method to prepare individual atoms is therefore non-deterministic, with a filling probability of one tweezers of $p\sim 0.5$. This makes its extension to large arrays of tweezers  (see Sec.~\ref{Sec:microtraps_arrays}) difficult: the probability to find a configuration where $N$ tweezers are filled at the same time decreases like $p^N$. This triggered investigations on how to improve the loading efficiency of optical tweezers. Two methods have been demonstrated so far. The first one, proposed and demonstrated by the Wisconsin group~\cite{Saffman2002,Ebert2014}, uses the Rydberg blockade in a small atomic ensemble trapped in a tight dipole trap and achieved a filling probability $p\simeq 0.62$. The second method demonstrated in Otago~\cite{Grunzweig2010,Carpentier2013,Fung2015} and at JILA~\cite{Lester2015} relies on a tailoring of the light-assisted collisions in the tweezers, and led to loading probabilities $p\sim0.90$. 

\subsection{Arrays of microtraps}\label{Sec:microtraps_arrays}

Once demonstrated the trapping of individual atoms in a trap, the next step in view of (scalable) quantum engineering applications is to create controlled arrays of such traps, each of them containing an individual atom. 

A first, natural approach consists in using optical lattices, i.e. periodic optical dipole potentials obtained by interfering several laser beams. One can use large-period optical lattices (with a lattice spacing on the order of a few microns, obtained by using interfering beams that make a small angle with each other), and load in a sparse way single atoms in the resulting array of  microtraps~\cite{Nelson2007}. Single-site imaging is relatively easy for such large-period lattices, and coherent single-site manipulations of individual atoms in such settings can also be achieved, even in 3D settings~\cite{Wang2015}. Another approach consists in using usual, short-period ($\sim 500$~nm) optical lattices, and loading ultracold atoms in a single 2D plane. There, single-site resolution requires the use of advanced high numerical aperture objectives, realizing a so-called \emph{quantum gas microscope}~\cite{Bakr2009,Sherson2010}. One of the assets of such an approach, despite its high technical complexity, is the possibility to use a Mott insulator to achieve single-atom filling with probabilities in excess of 90\% per site. Single-atom addressing, using techniques developed in the context of 3D optical lattices~\cite{Lee2007,Lee2013}, can also be achieved in those settings~\cite{Weitenberg2011}. A drawback of this latter approach is that for a large variety of Rydberg experiments, small lattice constants limit the range of coupling strengths that one can use.

\begin{figure}[t]
\centering
\includegraphics[width=8.5cm]{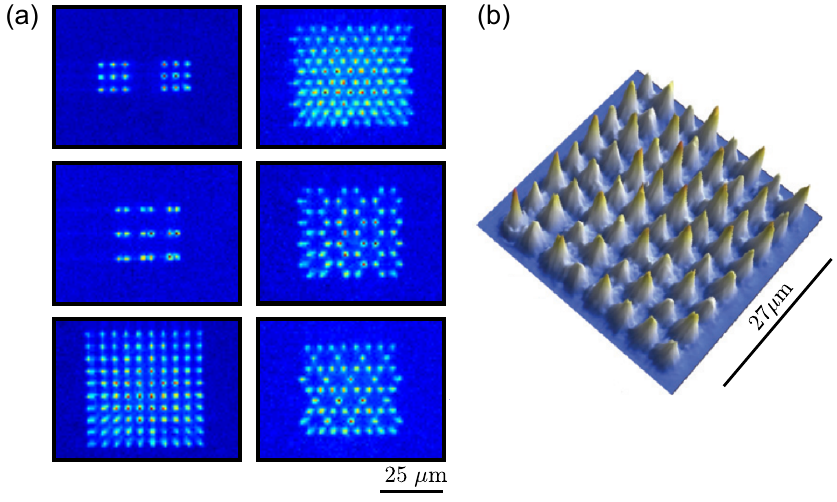}
\caption{Generating arrays of microtraps. (a) Averaged fluorescence images of single atoms trapped in microtrap arrays generated using a spatial light modulator. (b) Array of $8\times8$ blue-detuned Gaussian beams created by diffractive optical elements, resulting in $7\times7$ trapping sites. Figure adapted from Refs.~\cite{Nogrette2013,Piotrowicz2013}.}
\label{fig:trap_arrays}
\end{figure}

A second approach, which allows for more flexible geometries, consists in optically creating several `copies' of an optical tweezers, thus realizing arrays of microtraps. For this, one can use holographic methods~\cite{Bergamini2004,Nogrette2013} (see Fig.~\ref{fig:trap_arrays}a), diffractive optics~\cite{Knoernschild2010}, or microfabricated optical elements~\cite{Dumke2002}.
Holographic optical tweezers offer a versatile solution regarding accessible geometries. Using a programmable spatial light modulator to imprint an arbitrary phase on a beam prior focusing, it is possible to replicate a single optical tweezers into hundreds of traps in arbitrary geometries~\cite{Bergamini2004,Nogrette2013}. Fast, single-site addressing can be achieved by adding an extra beam controlled by acousto-optic deflectors to add light-shifts on targeted sites~\cite{Labuhn2014}. The number of traps can be massively increased, at the expense of some flexibility on geometry, by making use of microfabrication techniques, as pioneered by the group of G. Birkl~\cite{Dumke2002}. More than $10^4$ high numerical aperture micro lenses can be fitted on an area of 1 mm$^2$, while allowing for micrometer size traps~\cite{Schlosser2011}. 

For experiments that rely on Rydberg excitation, however, red-detuned traps have some limitations. The trapping light reduces the lifetime of the Rydberg stated via photoionization, and produces position-dependent differential light shifts between ground and Rydberg states. To avoid these problems, the traps are generally turned off during Rydberg excitation, increasing atom losses. A solution to this problem has been investigated by Saffman and coworkers~\cite{Piotrowicz2013,Zhang2011}. They showed that for certain Rydberg states, it is possible to find trap wavelengths (called quasimagic wavelengths) for which the ground and excited states are shifted by the same amount. For alkali atoms, it implies the need for blue-detuned light, which they use to create 2D arrays of dark traps by weakly overlapping Gaussian beams (see Fig. \ref{fig:trap_arrays}b). The geometry of the obtained arrays is however more constrained than in the case of arrays of red-detuned optical tweezers.

\subsection{Laser excitation to Rydberg states}

For alkali atoms, optical transitions between a given ground-state  and Rydberg states with principal quantum numbers $n = 40-200$ lie in the UV domain, with wavelengths in the range $230-320$ nm. Direct, coherent optical excitation with CW lasers has recently been demonstrated for single cesium atoms~\cite{Hankin2014}, requiring a powerful UV laser. The Rabi oscillations between the states $|6S_{1/2},F=4,m_F=4\rangle$ (prepared by optical pumping)  and $|84P_{3/2},m_j=3/2\rangle$ had a frequency $\sim 1$ MHz. Note that single-photon transitions do not allow to cancel the Doppler effect, and that, due to electric dipole selection rules ($\Delta L = \pm 1$), such schemes limit Rydberg excitation to P-states. 

Most of the experiments using individual alkali atoms rely instead on two-photon transitions. In rubidium, the most frequently used scheme is the combination of 795 (780) nm and 474 (480) nm photons off-resonant from the intermediate state $5 P_{1/2}$ ($5 P_{3/2}$)~\cite{Johnson2008,miroshnychenko2010,Zuo2009,Loew2013}. The ``inverted'' scheme $5S-6P-nS/nD$ with 420 and 1016 nm light is also possible, but has not been used so far with individual rubidium atoms. This inverted scheme ($6S-7P-nS/nD$) was implemented with individual cesium atoms combining 459 and 1038 nm lasers~\cite{Maller2015}. 

\begin{figure}[t]
\centering
\includegraphics[width=8.5cm]{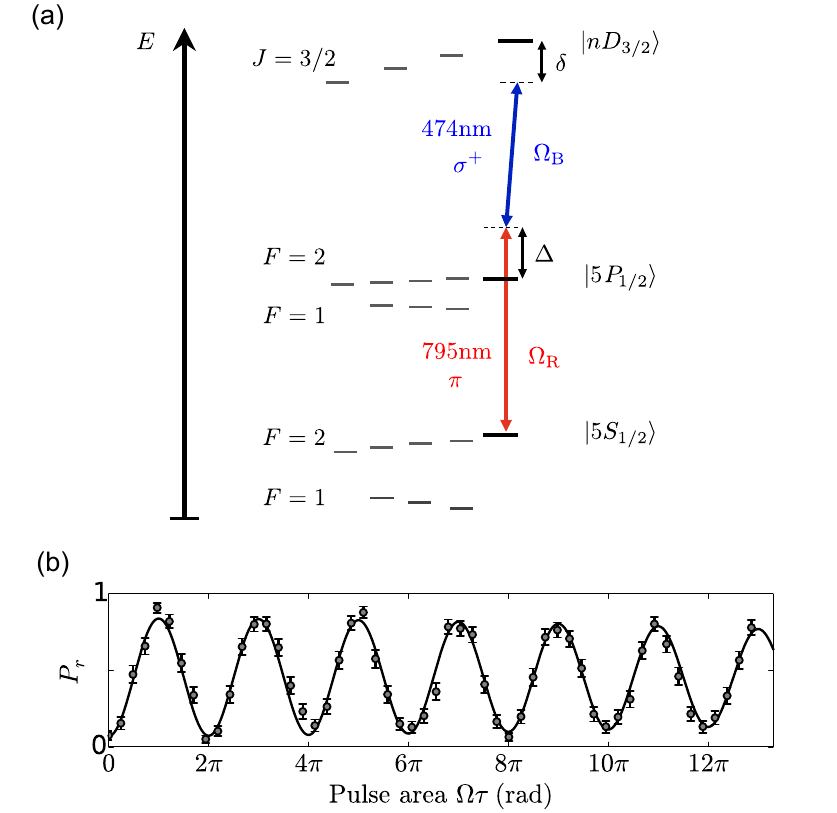}
\caption{(a) Two-photon excitation scheme to $nD_{3/2}$ Rydberg states in Rb used at Institut d'Optique. A $\pi$ polarized 795 nm light field couples the ground state $\ket{g} = \ket{5 S_{1/2}, F=2, m_F=2}$ prepared by optical pumping with the intermediate state $\ket{5 P_{1/2}, F=2, m_F=2}$ with a detuning $\Delta = 2 \pi \times 740$ MHz. In the second excitation step a $\sigma^+$-polarized 474 nm beam populates the Rydberg state $\ket{r}= \ket{nD_{3/2},m_j=3/2}$. (b) Typical single-atom Rabi oscillations between the ground $\ket{g}$ and Rydberg state $\ket{r}$.}
\label{fig:excitation}
\end{figure}

In the limit of a large detuning $\Delta$ with respect to the intermediate state, 
$\Delta \gg \Omega_R,\Omega_B$ (with $\Omega_R$, $\Omega_B$ the red and blue Rabi frequencies, respectively), 
the three-level system shown in Fig.~\ref{fig:excitation}a can be reduced to an equivalent two-level system, where the ground state $\ket{g}$ couples to the Rydberg state $\ket{r}$ with an effective Rabi frequency $\Omega_{\rm eff}$ and an effective detuning $\delta_{\rm eff}$ given by:
\begin{equation}
\Omega_{\rm eff} = \frac{\Omega_R \Omega_B}{2 \Delta}\,\, \mbox{ and } \,\,\,\,
\delta_{\rm eff} = \delta- \left( \frac{|\Omega_R|^2}{4 \Delta} - \frac{|\Omega_B|^2}{4 \Delta}\right).
\label{eq:Rabi_effective}
\end{equation}

If this condition does not hold, spontaneous emission via the intermediate state is not negligible on the excitation timescale and leads to a loss of coherence in the excitation. The spontaneous scattering rate $\Gamma_{\rm eff}$ can be estimated perturbatively from the average population in the intermediate state and its decay rate $\Gamma$ as:

\begin{equation}
\Gamma_{\rm eff} = \Gamma\left(\frac{\Omega_R^2 +\Omega_B^2}{4 \Delta^2}\right).
\label{eq:scattering_rate}
\end{equation}

In addition to a low scattering rate, coherent coupling between the ground state and the Rydberg state requires effective Rabi frequencies higher than the linewidth of the Rydberg state (e.g. $\sim 2\pi\times 1.5$ kHz for $62 D_{3/2}$ in Rb), and sufficiently narrow laser linewidths. As an example, in the experiment at Institut d'Optique, typical effective Rabi frequencies $\Omega_{\rm eff} \sim 2\pi \times 1-10 $ MHz can be obtained with between 100 $\mu$W and 10 mW of laser power at 795 nm (focused to a beam waist of 120 $\mu$m) and $\sim$ 100 mW at 474 nm  (for a beam waist of 20 $\mu$m). Both excitation lasers are frequency locked to an ultra-stable, high-finesse ULE cavity ($\mathcal{F} > 20000$), providing overall laser linewidths $<$ 10 kHz.  With this setup we routinely obtain Rabi oscillations with small damping rates and visibilities exceeding 90\% (see Fig.~\ref{fig:excitation}b). Using similar techniques, comparable Rabi frequencies and visibilities are obtained by the Wisconsin group, either using cesium or rubidium (see also the work at Chofu university~\cite{Zuo2009}).

\subsection{Electric fields}\label{sec:electric_fields}

The huge polarizability $\alpha\propto n^7$ of Rydberg atoms, arising from their large transition dipole moments, makes them very sensitive to electric fields. For the Rydberg states of alkali atoms accessible by laser excitation from the ground state, the angular momentum is low ($l\lesssim 3$) and, as a consequence of the quantum defects, the  Stark effect is quadratic in low electric fields. As an example, for a rubidium atom, a residual electric field of $\sim 150 {\rm mV}/{\rm cm}$ is enough to shift the $\ket{59 D_{3/2}, M_J=3/2}$ state by 4.5 MHz. It is therefore important to accurately control the electrostatic environment of the atoms to prevent unwanted shifts. In experiments stray electric fields are reduced by grounding most of the surfaces surrounding the atoms. This includes the aspheric lenses used to focus the tweezers beam at Institut d'Optique and by the Sandia National Laboratory group, located $\sim 2-10$ mm away from the plane of the atoms, that are coated with a 200 nm thin layer of indium tin oxide (ITO)\footnote{However, as noted in~\cite{Hankin2014}, UV light can produce surface charging of the ITO layer close to the atoms via the photoelectric effect, with detrimental effects in the manipulation of Rydberg states.}. The Institut d'Optique team also cancels actively  any residual DC field in three directions by a set of eight electrodes in an octopole configuration that can be addressed independently~\cite{Loew2013}. The field plates are also used to apply finely controlled pulsed fields, to Stark-tune the Rydberg state energies. This allows, for example, switching on dipole-dipole interactions between the atoms at a F\"{o}rster resonance, by matching the resonance condition in a given time window, as will be shown in Section~\ref{Sec:exp_Forster}.

\subsection{Microwave manipulation in the Rydberg manifold}
\label{sec:mw}

Rydberg states are coupled to other Rydberg states by electric dipole transitions in the microwave (MW) domain. Due to the dipole moment between nearby states scaling as $n^2$, even a small amount of microwave power is enough to drive the transition with a high Rabi frequency. This feature has established Rydberg atoms as very sensitive probes with subwavelenth resolution that can be potentially used as calibration standards in MW-imaging~\cite{Sedlacek2012,Holloway2014}, and it is also a very convenient tool for the manipulation of Rydberg states. From an optically excited Rydberg state, other nearby Rydberg states can be accessed with moderate MW power. 

\begin{figure}[t]
\centering
\includegraphics[width=8.5cm]{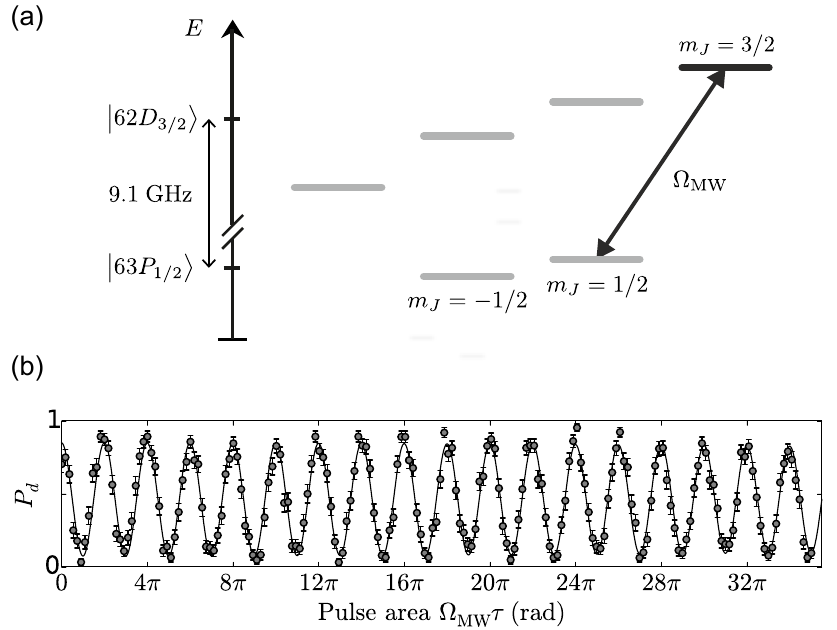}
\caption{(a) A microwave field couples the $62 D_{3/2}$ and $63 P_{1/2}$ Rydberg states in Rb. The MW polarization is a combination of $\sigma_+$ and $\sigma_-$ polarizations at the position of the atoms. A 6.6 G magnetic field shifts the Zeeman sub-levels so that only two levels remain resonant with the MW field. (b) Rabi oscillation between the two Rydberg states. Figure adapted from~\cite{Barredo2015}. }
\label{fig:excitationMW}
\end{figure}

The Institut d'Optique group has demonstrated the coherent coupling between the $\ket{62 D_{3/2}}$ and $\ket{63 P_{1/2}}$ Rydberg states, as shown in Fig.~\ref{fig:excitationMW}~\cite{Barredo2015}. A 9.1~GHz driving field is applied with a 5 mm electric dipole antenna placed outside the vacuum chamber, 20 cm away from an atom trapped in a tweezers. The transition dipole element between the two states, $\bra{62 D_{3/2}, m_j=3/2} \hat{d_+} \ket{63 P_{1/2}, m_j=1/2} \simeq 2858 e a_0$,  and $\sim 40 \mu$W of  microwave power are enough to drive Rabi oscillations at a frequency of $\Omega_{\rm MW} = 2 \pi \times 4.6$ MHz. Well contrasted oscillations, with almost no damping over several microseconds, are only observed if the underlying level structure resembles a two-level system. To achieve this, we apply a 6.6 G magnetic field that lifts the Zeeman degeneracy and ensures that only two levels are addressed with the MW field, even if its polarization at the position of the atoms is not well controlled (Fig.~\ref{fig:excitationMW}). 

\subsection{Detection of Rydberg states}

Positive detection of Rydberg states is generally accomplished via field ionization and subsequent detection of the electron/ions with over 90\% efficiency~\cite{Gallagher1994,Loew2013}. This method has been traditionally used for Rydberg detection in cold atom clouds. For the separations of a few microns attained in optical tweezers based setups, site-selective detection via multichannel plates is challenging, and would probably require the use of a tip imaging probe close to the atoms~\cite{Schwarzkopf2013}. 

Single-atom trapping in arrays of optical tweezers, however, naturally provides another detection method based on atom losses. The dipole trap laser operating around 850-950 nm induces a small positive light shift of $\sim 1$ MHz (for a 20~MHz trap depth for ground-state atoms) for Rydberg states with principal quantum numbers $n>50$. Therefore, Rydberg atoms cannot be trapped in the dipole traps, and due to their finite temperature,  they have ample time to escape the trapping region within their lifetime. For typical experimental parameters used by the Wisconsin, Sandia and Institut d'Optique groups, the probability for a Rydberg atom to remain in the trapping region after 50 $\mu$s is below 10\%. 
This detection method therefore maps an excitation to a Rydberg state onto a loss of the atoms following the excitation. As an example, the Institut d'Optique reported an efficiency of this method of 97\%~\cite{Barredo2014}, which means that in only 3\% of the cases the loss of atoms is not due to excitation to a Rydberg state. 

This detection technique can be made Rydberg-state-dependent, allowing to discriminate between states with different parities. This was illustrated in~\cite{Barredo2015}, by combining microwave and optical excitation (see Fig.~\ref{fig:excitationMW}): after excitation to an $nD$ state, the microwave pulse transfers part of the population to a nearby $n'P$ state. The remaining fraction $nD$ is mapped down to the ground state, where its presence is inferred by a fluorescence measurement. An atom loss is now the signature of a transfer of the atom to the $nP$ state, which is not coupled back to the ground state.

The obvious drawback of this method is that any unwanted loss (e.g. collisions with the background gas...) mimics a Rydberg excitation. A way around this consists in removing all ground state atoms with a resonant laser pulse while the other atoms are in a Rydberg state, before de-exciting the Rydberg atoms via stimulated emission to the intermediate state and detecting the fluorescence. The Rydberg detection now  relies on a positive detection. This method was implemented for atoms in optical lattices~\cite{Schauss2012}, with a detection efficiency limited to $\sim 80\%$ so far. 

\section{Demonstration of the Rydberg blockade and entanglement with two atoms}\label{Sec:blockade_exp}

The experimental effort aiming at observing the Rydberg blockade between two atoms 
started at the University of Wisconsin shortly after the initial proposals. It was backed by an in-depth theoretical analysis in 2005~\cite{Saffman2005}. The Institut d'Optique team started in 2008. Both  groups observed the blockade in 2008 and used it to demonstrate in 2009 the entanglement of two atoms (Institut d'Optique) and a CNOT gate (Wisconsin). 
The group at Sandia National Laboratory joined the effort a few years later and was able to observe 
the blockade and to entangle two atoms using a dressed Rydberg interaction. 

\subsection{Rydberg blockade: the Wisconsin experiment~\cite{Urban2009}}

\begin{figure}[t]
\begin{center}
\includegraphics[width=6.8cm]{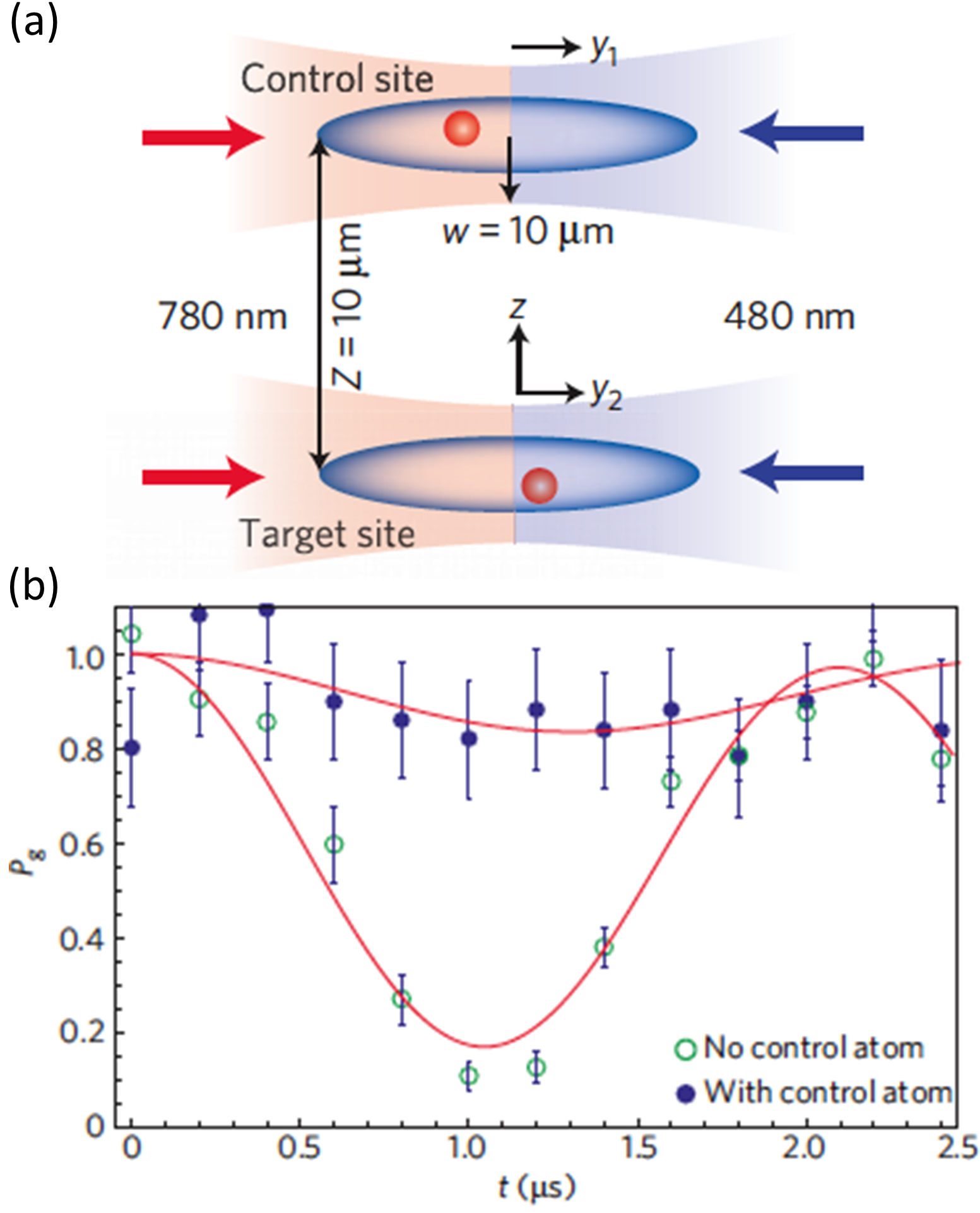}
\caption{Observation of the blockade between two atoms by the group at University of Wisconsin. (a) The  two traps are separated by $R=10\ \mu$m and the two atoms can be excited separately from each other by independent laser beams. (b) Rabi oscillation on the target atom without or with the control atom in the Rydberg state. Here $P_g$ is the probability that the target atom is still in the ground state at the end of the laser excitation to the Rydberg state.  Each data point is an average over many realizations of the experiment in order to measure the probabilities.
Figures from~\cite{Urban2009}.}\label{fig:Blockade_Saffman}
\end{center}
\end{figure}

In this experiment, the group of M. Saffman at the University of Wisconsin (USA)  trapped two rubidium atoms in two dipole traps  separated by a distance of $R\approx 10$~$\mu$m, see figure~\ref{fig:Blockade_Saffman}(a). Here, each atom can be excited independently, one atom being considered as the control atom, the other one as the target atom. After preparing the atoms in  the hyperfine ground state $|g\rangle=|5S_{1/2},F=2,m_F=0\rangle$, the team first excited the target
atom to the Rydberg state $|r\rangle=|nD_{5/2},m_j=5/2\rangle$ ($n=79$ or 90) using a two-photon transition.
As they varied the duration of the excitation, they observed the characteristic 
Rabi oscillations between the states $|g\rangle$ and $|r\rangle$ (see Fig.~\ref{fig:Blockade_Saffman}b).

In order to demonstrate the blockade, the authors started by exciting the control atom to the Rydberg state by applying a $\pi$ pulse. They then sent the excitation laser on the target atom and observed that the probability to excite it to the Rydberg state was strongly suppressed as shown in figure~\ref{fig:Blockade_Saffman}(b). This is the signature of the Rydberg blockade and  the proof that the Rydberg excitation of the target atom is
controlled by the state of the control atom. Note that in this addressable version of the Rydberg blockade, the atoms are not entangled at the end of the sequence.

\subsection{Rydberg blockade: the Palaiseau experiment~\cite{Gaetan2009}}

\begin{figure}[t]
\begin{center}
\includegraphics[width=7.5cm]{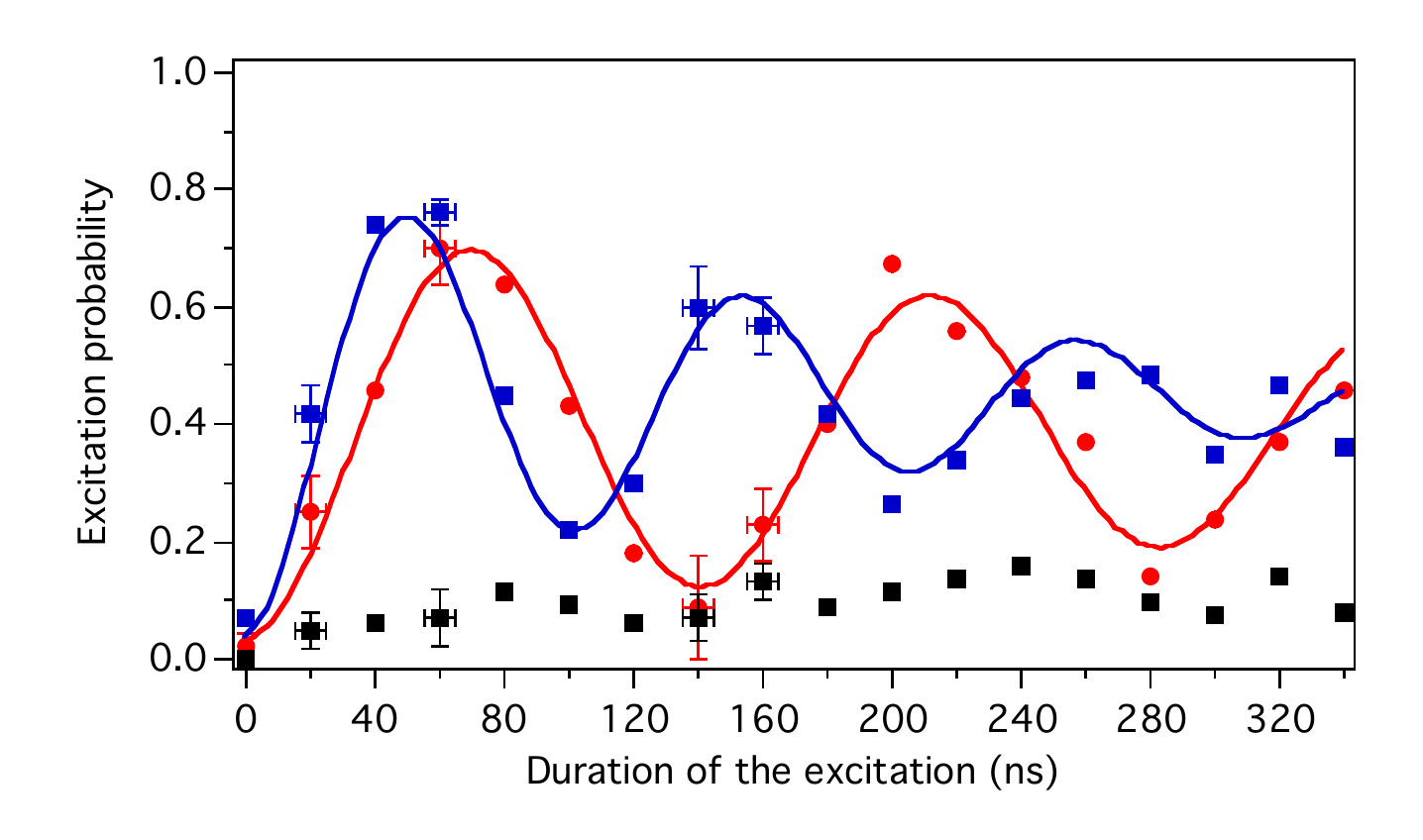}
\caption{Observation of the blockade and of the collective excitation of two atoms by the Institut d'Optique group. Here the two  traps are separated  by 4 $\mu$m.  The excitation lasers do not address the atoms  independently, as their size is much larger than  the interatomic distance. The atoms are excited by a two-photon transition. Red disks: probability to excite atom A  alone when the trap B is empty. Black squares: probability to excite the two atoms.  Blue squares: probability to excite one and only one of the two atoms.  Each data point is an average of 100 realizations of the experiment, in order to  calculate the probabilities. Figure from~\cite{Gaetan2009}.}\label{fig:Blockade_IO}
\end{center}
\end{figure}

In the Institut d'Optique experiment, the two atoms $A$ and $B$ were trapped in two dipole traps separated by a distance of 4~$\mu$m.  There, the Rydberg excitation laser does not address a specific atom.  

The group first measured a Rabi oscillation with only atom $A$ present, the  second trap being empty: it observed the Rabi oscillations between state  $|g\rangle=|5S_{1/2},F=2,m_F=2\rangle$ and $|r\rangle=|58D_{3/2},m_j=3/2\rangle$, as shown by the red disks in Fig.~\ref{fig:Blockade_IO}. The experiment is then repeated when the two traps contain one atom each. The group measured the probability  $P_{rr}$ to excite the two atoms (black squares) and the probability $P_{rg}+P_{gr}$ to excite only one of the two atoms (blue squares). The  probability of exciting the two atoms at the same time is indeed suppressed, as it should be for two atoms in the blockade regime.  However, the probability to excite one of the two atoms does oscillate,  and the oscillation frequency is larger than when only one atom is present. The ratio of the two measured Rabi frequencies is 1.38, in very good agreement with the expected $\sqrt{2}$ factor.  This enhancement of the oscillation frequency is the signature of the collective excitation of the two atoms: in the blockade regime, the laser couples the two collective states $|gg\rangle$ and $|\psi_+(\phi)\rangle=(|rg\rangle+e^{i\phi}|gr\rangle)/\sqrt{2}$. Here the phase factor $\phi= ({\boldsymbol k}_{\rm R}+{\boldsymbol k}_{\rm B})\cdot({\boldsymbol r}_A-{\boldsymbol r}_B)$ is imposed by the geometry of the red and blue excitation lasers (wave vectors ${\boldsymbol k}_{\rm R}$ and ${\boldsymbol k}_{\rm B}$, respectively) and the positions ${\boldsymbol r}_A$, and ${\boldsymbol r}_B$ of the atoms. The positions of the atoms are fixed during a single realization of the experiment but vary from one realization to another, leading to a shot-to-shot variation of the phase $\phi$ by more than $2\pi$. Strictly speaking, the experiment therefore produces a statistical mixture of states  $|\psi_+(\phi)\rangle$.

\subsection{Rydberg blockade: the Sandia experiment~\cite{Hankin2014}}

The team at Sandia National Laboratory used two cesium atoms trapped in two tweezers separated by 6.6~$\mu$m.  The Rydberg excitation connects the hyperfine ground state $|g\rangle=|6S_{1/2},F=4,m_F=0\rangle$  to the Rydberg state $|r\rangle=|84P_{3/2},m_j=3/2\rangle$ by a single step process at 319 nm. The laser does not address a specific atom and the group observed the  same signatures as in the Palaiseau experiment: a suppression of the probability to excite the  two atoms at the same time and the $\sqrt{2}$ enhancement of the collective Rabi oscillation. 

\subsection{Demonstration of a CNOT gate and entanglement between two atoms}\label{Sec:exp_entanglement_CNOT}

The immediate step after the demonstration of the Rydberg blockade for the three groups was the demonstration  of entangling operations. The three groups followed three different approaches. 

The Institut d'Optique group started from the collective state $|\psi_+(\phi)\rangle=(|rg\rangle+e^{i\phi}|gr\rangle)/\sqrt{2}$ produced as a consequence of the blockade 
and mapped the Rydberg state $|r\rangle$ to the hyperfine ground state  $|g'\rangle=|5S_{1/2},F=1,m_F=1\rangle$ using a second red laser close to 795 nm. In doing so, the phase factor $\phi$ is erased~\cite{Wilk2010}, provided the atoms do not move during the excitation and mapping pulses. The final state should then be close to the 
Bell state $|\Psi_+\rangle= (|g'g\rangle+|gg'\rangle)/\sqrt{2}$. This state also presents the advantage of being trapped in the tweezers and long-lived. The team measured the fidelity $|\langle \Psi_{+}|\Psi_{\rm exp}\rangle|^2$ of the state prepared in the experiment $|\Psi_{\rm exp}\rangle$ by applying a global rotation using Raman lasers coupling the two states $|g\rangle$ and $|g'\rangle$. They could extract two types of fidelities. The first one corresponds to the fidelity with which the state $|\Psi_+\rangle$ is prepared in the experiment, and amounts to 0.46. However, there is a 61\% probability to lose at least one of the two atoms during the entangling sequence, which leads to a fidelity of the remaining pairs of 0.75, larger than the 0.5 threshold to claim entanglement~\cite{Sackett2000}. A detailed analysis of the experiment was performed in~\cite{Gaetan2010}. 

\begin{figure}[t]
\begin{center}
\includegraphics[width=6.5cm]{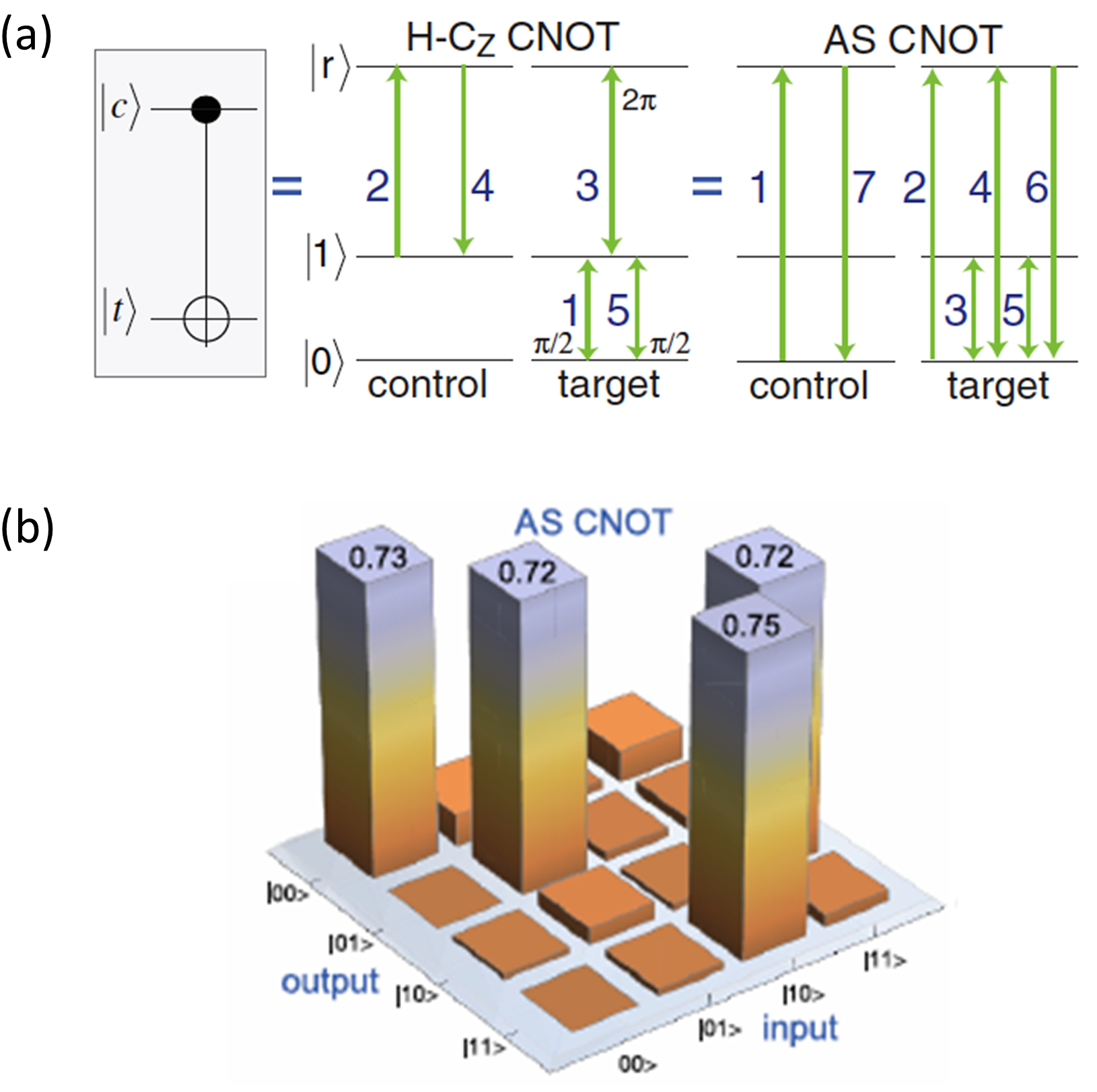}
\caption{ Demonstration of a CNOT gate by the University of Wisconsin group. (a) Sequences of pulses used  to implement the gates. The two sequences lead to a CNOT gate. (b) Experimental truth table  for the  AS CNOT sequence.  Figures from~\cite{Isenhower2010}.}\label{fig:CNOT_Wisconsin}
\end{center}
\end{figure}

The Wisconsin group demonstrated a CNOT gate~\cite{Isenhower2010}, thanks to their ability to perform local addressing of each atom. To do so, they used two types of sequences (involving respectively 5 and 7 pulses, see Fig.~\ref{fig:CNOT_Wisconsin}a) to implement a variant  of the original proposal~\cite{Jaksch2000}.  This two-bit gate involves two hyperfine ground states of the rubidium atom, labeled $|0\rangle= |5S_{1/2},F=1,m_F=0\rangle$ and $|1\rangle= |5S_{1/2},F=2,m_F=0\rangle$, and the Rydberg state $|r\rangle= |97D_{5/2}, m_j=5/2\rangle$ as an intermediate state in the sequence. The Rydberg blockade is the underlying mechanism, which allows or not the flipping of the state of the target atom depending on the state of the control atom. The Wisconsin group reported a fidelity of the gate truth table of 0.73~\cite{Isenhower2010} (Fig.~\ref{fig:CNOT_Wisconsin}b).  

They also used the gate to demonstrate the preparation of the four entangled Bell states. Starting from the control atom prepared in the superposition $\left(|0\rangle +|1\rangle\right)/\sqrt{2}$, and the target state in $|0\rangle$, the action of the CNOT gate leads to the final two-atom state $\left(|00\rangle +|11\rangle\right)/\sqrt{2}$. The fidelity of the entangled states was reported to be around~0.48. As in the Palaiseau experiment, atom losses during the sequence lead to a probability of having the two atoms at the 
end of the sequence of 83\%. The corrected fidelity is therefore 0.58. A few months later, the group reported an improved fidelity of 0.58 of the entanglement without accounting for the loss and 0.71 when correcting for the atom losses~\cite{Zhang2010}. The non-corrected loss is therefore already higher than the threshold for entanglement at 0.5. Finally, the group recently implemented the original proposal of~\cite{Jaksch2000} between two next-nearest-neighbor cesium atoms trapped in an array of 49 traps separated by 3.6~$\mu$m. The fidelity of preparation of the $(|00\rangle+|11\rangle)/\sqrt{2}$ state is 0.73 including the losses and 0.79 after correction~\cite{Maller2015}. 

\subsection{Demonstration of two-atom entanglement using a dressed Rydberg interaction~\cite{Jau2015}}

The Sandia team also used the Rydberg blockade to entangle two atoms. However they did it while keeping the atoms in their hyperfine ground states, contrarily to the Institut d'Optique experiment. The protocol uses a Rydberg-dressed interaction proposed initially by I. Bouchoule and K. Moelmer~\cite{Bouchoule2002} in 2002, further expanded by G. Pupillo and co-workers~\cite{Pupillo2010} in 2010. The principle of the Rydberg-dressed interaction is the following~\cite{Johnson2010}: a laser couples the ground state $|g\rangle$ to the Rydberg state $|r\rangle$, with a Rabi frequency $\Omega$ and a detuning $\Delta$. This laser admixes the two atomic states, giving to the ground state a part of the Rydberg characteristics, therefore allowing two atoms in the ground state to interact.

\begin{figure}[t]
\begin{center}
\includegraphics[width=7.5cm]{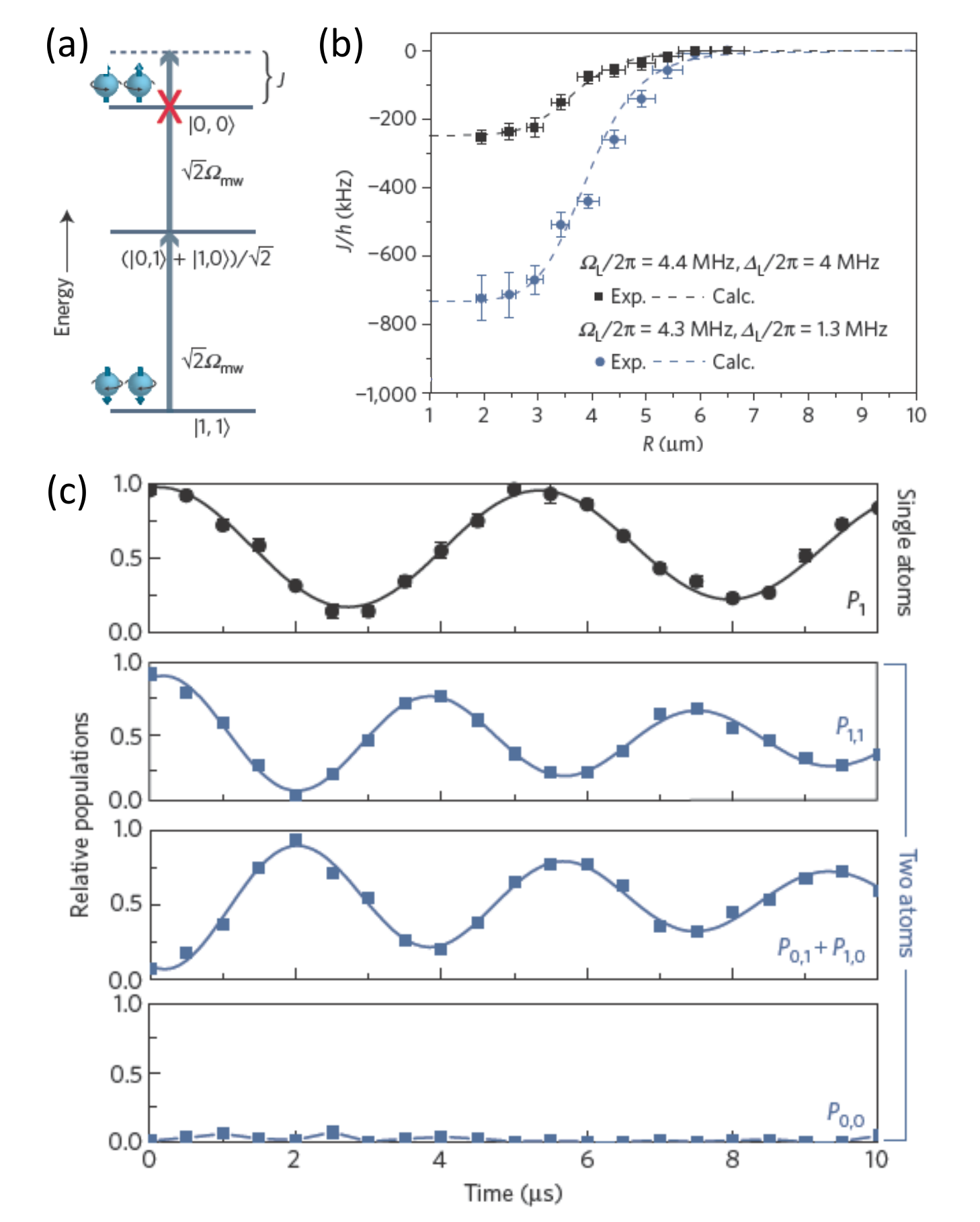}
\caption{ Demonstration of a dressed Rydberg interaction by the Sandia National Laboratory group. 
(a) Two-atom spectrum in the presence of the dressing laser coupling the ground state $|0\rangle$ to the Rydberg state $|r\rangle= |64P_{3/2}, m_j=3/2\rangle $. (b) Measurement of the dressed interaction for two different sets of ($\Omega,\Delta$). (c) Collective Rabi oscillations between states $|11\rangle$ and $(|01\rangle+|10\rangle)/\sqrt{2}$.  The blockade is reflected by the negligible population $P_{00}$. The upper curve ($P_1$) is the Rabi oscillation between states $|0\rangle$ and $|1\rangle$ when only one atom  is used. Figures from~\cite{Jau2015}.}
\label{fig:dressed_sandia}
\end{center}
\end{figure}

It can be shown that for two atoms located within a  blockade radius $R_{\rm b}$,  the effect of the dressing is to shift the two-atom ground state $|gg\rangle$ by an amount $J= {\hbar \over 2}[\Delta+{\rm sign}(\Delta)(\sqrt{\Delta^2+2\Omega^2}-\sqrt{\Delta^2+\Omega^2})]$, which is  independent of the inter-atomic distance $r$ as long as $r< R_{\rm b}$. Applying this idea to an atom with two hyperfine states $|0\rangle$ and $|1\rangle$ with the state $|0\rangle $ coupled to the Rydberg state $|r\rangle$, the two-atom spectrum restricted to the basis $\{|00\rangle, (|01\rangle+|10\rangle)/ \sqrt{2},|11\rangle\}$ is anharmonic (see Figure~\ref{fig:dressed_sandia}a). A pair of Raman lasers (or a microwave field) tuned to the  $|0\rangle-|1\rangle$ transition cannot excite two atoms initially in state $|00\rangle$ to the state $|11\rangle$. This is the exact equivalent of the blockade experiment at Institut d'Optique,  but in the ground state manifold. The Sandia group implemented this idea by using two moving traps, each containing  one cesium atom. They initially prepared each atom in the state $|0\rangle= |6S_{1/2},F=4,m_F=0\rangle$, while separated by a distance of 6.6~$\mu$m. Then, they approach the two atoms at a distance of 3~$\mu$m to  enhance their interaction, while applying a pair of Raman beams to drive the $\ket{0}-\ket{1}$ transition with the dressing beam at 319 nm on at the same time. By scanning the frequency of the Raman laser, they could measure the dressed interaction energy (see Figure~\ref{fig:dressed_sandia}b). Working in the Rydberg blockade configuration (Raman laser tuned on resonance with the $0-1$ transition) they could observe the characteristic $\sqrt{2}$ enhancement of the Rabi frequency (see Figure~\ref{fig:dressed_sandia}c) and generate the entangled state $(|01\rangle+|10\rangle)/\sqrt{2}$ with a fidelity of  0.81. However, there is still a 40\% probability to lose at least one atom during the sequence. 

This experiment is the first demonstration of Rydberg-dressed interactions. Key to the success was the use of a single-step excitation at 319 nm, which is not plagued by the spontaneous emission from an intermediate level, as is the case for two-photon excitation. The group also analyzed a scheme for implementing a controlled-Z gate using this approach~\cite{Keating2015}.

\subsection{Conclusion on the blockade and entanglement experiments}

In the early demonstrations at Wisconsin, Institut d'Optique and Sandia, the quality of the blockade was not perfect, usually featuring probabilities of double excitation as high as 15-20\%. This plagued the fidelities of the entangled states prepared and of the CNOT gate. Detailed theoretical investigations of the measured fidelities~\cite{Zhang2012,Maller2015} seem to indicate that the limitations are mainly technical, and therefore could be overcome. This triggered the construction of a new generation of dedicated experimental setups, including in particular control of the electric fields. These experiments are starting to produce results, and excellent blockade with double excitation probabilities as low as a few percent has been observed for two and three atoms (see Section~\ref{Sec:large_atom_numbers}). At the moment, the quality of the blockade should not be the main limitation in the fidelity of entangling operations.

\section{Measurement of the interaction energy between two Rydberg atoms}

\begin{figure}[t]
\centering
\includegraphics[width=8.5cm]{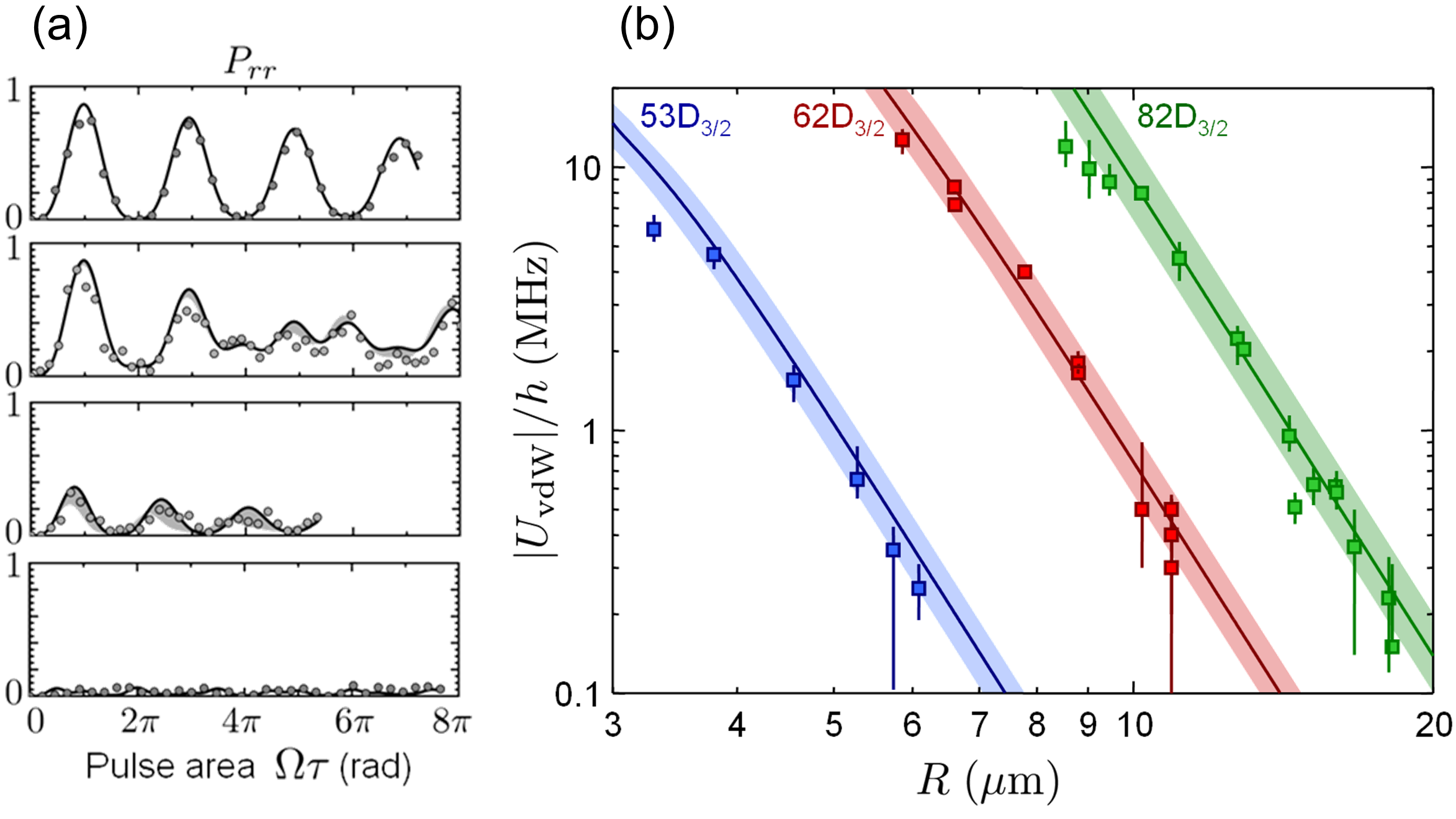}
\caption{Measurement of the van der Waals interaction between two Rydberg atoms. (a) Probability $P_{\rm rr}$ of exciting the two atoms to the Rydberg state $62D_{3/2}$, as a function of the area $\Omega\tau$ of the excitation pulse, for decreasing distances $R$ between the atoms (from top to bottom, $R=15,\;10,\;8.8,\;4\;\mu{\rm m}$). The lines are fits by the solution of a 4-level model with $U_{\rm vdW}$ as the only adjustable parameter. (b) The interaction obtained from such fits, as a function of $R$, for three different Rydberg states. The solid lines are the results of \emph{ab initio} calculations without any adjustable parameters (the shaded area represent uncertainty in the calibration of $R$). Figure adapted from~\cite{Beguin2013}. }
\label{fig:vdw_exp}
\end{figure}

In this section, we review a set of experiments, performed  between 2013 and 2015 at Institut d'Optique, on the measurement and control of the interaction between two individual $^{87}$Rb atoms held at well-defined positions, in the three regimes introduced in Section~\ref{Sec:theo_interaction}.

\subsection{Van der Waals interaction~\cite{Beguin2013}}

The basic idea to measure directly the interaction energy as a function of the distance $R$ between two atoms in $\ket{r}=\ket{nD_{3/2}}$ consists in working in the \emph{partial blockade} regime, i.e. when $\hbar \Omega \sim U_{\rm vdW}$. In this case, the dynamics of the system depends on both the Rabi frequency $\Omega$ and the interaction $U_{\rm vdW}$, which allows one to determine the latter.

The two atoms are initially prepared in the ground state, and then illuminated by Rydberg excitation lasers with Rabi frequency $\Omega$ for a time $\tau$. Figure~\ref{fig:vdw_exp}a shows the dynamics of the population of the doubly excited state $\ket{rr}$ (with $n=62$), for decreasing distances $R$ between the atoms. The top panel shows the almost not interacting case at large $R$, where ideally $P_{\rm rr}\simeq\sin^4(\Omega \tau/2)$ (the product of two independent Rabi oscillations). The bottom panel corresponds to a small enough $R$ such that the Rydberg blockade is effective and thus $P_{\rm rr}\simeq0$. For intermediate cases however, the dynamics is more involved, $P_{\rm rr}(\tau)$ showing a beating between incommensurate frequencies that depend on both $\Omega$ and $U_{\rm vdW}$. The solid lines are fits to the solution of the optical Bloch equations for the four-state system $\{\ket{gg},\ket{gr},\ket{rg},\ket{rr}\}$, where $U_{\rm vdW}$ is left as an adjustable parameter. 

Figure~\ref{fig:vdw_exp}b shows the obtained interaction energies, when the experiment is repeated for various distances $R$, and then for different principal quantum numbers $n$. One observes the $1/R^6$ scaling of the van der Waals interaction. The agreement with \emph{ab initio} calculations of the interaction (solid lines) is very good.

The same technique was subsequently used in~\cite{Barredo2014} to measure the angular dependence of the van der Waals interaction between two $nS_{1/2}$ or $nD_{3/2}$ states (see Figure~\ref{fig:angle}a,b). While in the first case the interaction is isotropic, the van der Waals interaction between $D$-states shows a clear anisotropy, varying by a factor $\sim 3$ when the angle between the quantization axis and the internuclear axis varies from $\theta=0$ to $\theta=\pi/2$. 

\subsection{F\"orster resonance~\cite{Ravets2014}}\label{Sec:exp_Forster}

\begin{figure}[t]
\centering
\includegraphics[width=8.5cm]{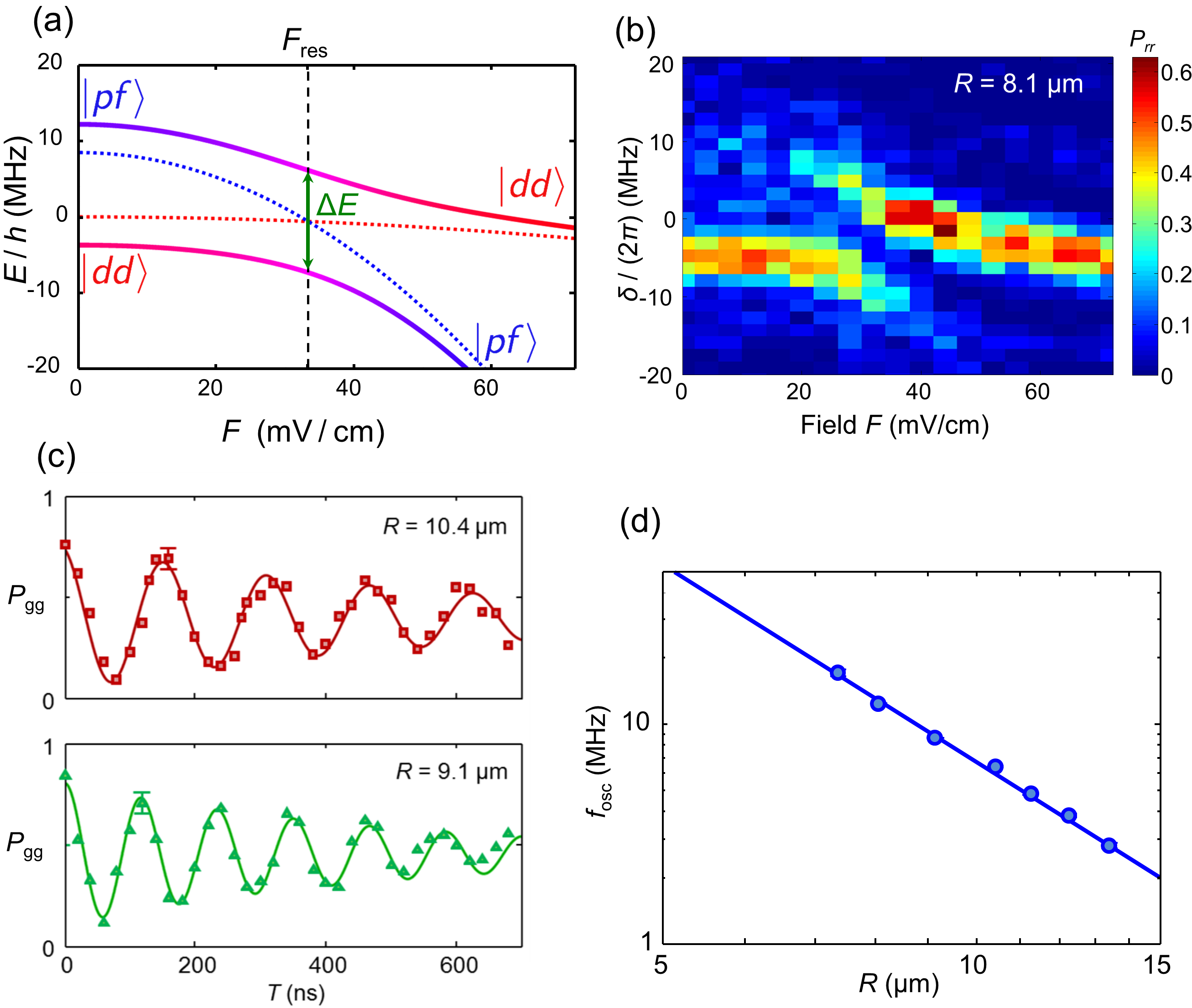}
\caption{Observation of a F\"orster resonance with two atoms. (a) Calculated Stark map of the two pair states $\ket{dd}$ and $\ket{pf}$ (see text), in the absence of dipolar coupling (dotted lines), and when the coupling is included (solid lines), giving an avoided crossing. (b) Experimental observation of the avoided crossing by laser spectroscopy.  (c) Oscillations in the probability for the atom pair to be in back to the state $\ket{dd}$ after staying for a time $T$ at resonance, for two distances $R$ between the atoms. Solid lines are fits by damped sines to extract the oscillation frequency. (d): Variation of the fitted oscillation frequencies with the distance $R$. The solid line is a fit by a power law, giving an exponent $-3.0(1)$. Figure adapted from~\cite{Ravets2014}.}
\label{fig:foerster_exp}
\end{figure}

In~\cite{Ravets2014}, we used our ability to apply arbitrary electric fields with electrodes to tune the pair state
$\ket{dd}=\ket{59D_{3/2},59D_{3/2}}$ on resonance with $\ket{pf}=\ket{61P_{1/2},57F_{5/2}}$. First, we performed a spectroscopic measurement to determine the exact value of the electric field giving rise to the avoided crossing between the two pair states (Figure \ref{fig:foerster_exp}a,b). In a second step, we studied the interaction in the time domain, by preparing first the system in $\ket{dd}$, and then switching abruptly (with a risetime of about 10~ns) the electric field to resonance, for an adjustable time $T$. A final optical readout pulse then allowed to measure the probability for the pair of atoms to be in $\ket{dd}$, showing coherent oscillations between the two coupled pair states (Figure \ref{fig:foerster_exp}c). The frequency of these oscillations scales as $1/R^3$ with the distance between the atoms. As compared to earlier studies of F\"orster resonances in disordered ensembles comprising a large number of atoms (see~\cite{Comparat2010} and references therein), this clean system consisting of only two atoms at controlled positions makes it possible to study directly the spatial dependence of the interaction and to observe its coherent character. 

A subsequent experiment~\cite{Ravets2015} measured the angular dependence of the dipolar interaction at resonance,  
observing the characteristic variation $1-3\cos^2\theta$ of the interaction with the angle $\theta$ between the internuclear axis and the quantization axis (see Fig.~\ref{fig:angle}c).

\subsection{Resonant dipole-dipole interaction~\cite{Barredo2015}}

In order to observe the resonant dipole-dipole interaction described in Section~\ref{Sec:theo_reson_dipdip}, a system of two atoms separated by a distance $R$ was prepared in the state $\ket{pd}$, where $\ket{p}=\ket{63P_{1/2},m_j=1/2}$ and $\ket{d}=\ket{62D_{3/2},m_j=3/2}$. This was achieved by (i) applying a light-shift on atom 1, using an addressing beam~\cite{Labuhn2014}, while using a global, two-photon, resonant Rydberg excitation pulse to bring atom 2 to $\ket{d}$; (ii) applying a microwave pulse at about 9.1~GHz (see section~\ref{sec:mw}) which brings atom 2 to $\ket{p}$; and (iii) exciting atom 1 to $\ket{d}$ with a resonant laser pulse (atom 2, in $\ket{p}$, is not affected by the Rydberg pulse).

\begin{figure}[t]
\centering
\includegraphics[width=8.5cm]{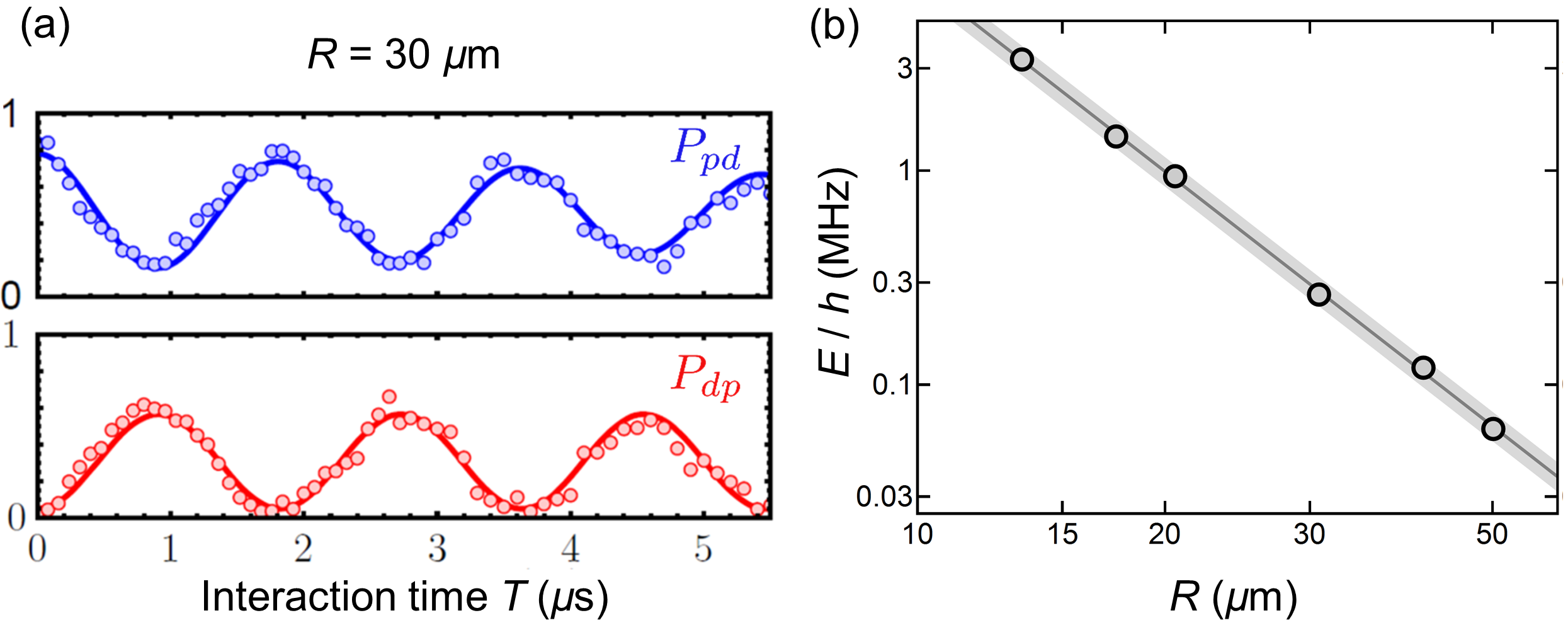} 
\caption{Resonant dipole-dipole exchange between two atoms. (a) Probabilities for the atom pair to be in $\ket{pd}$ and $\ket{dp}$ (see text) as a function of time, for two atoms separated by $R=30\;\mu{\rm m}$. (b) Oscillation frequency as a function of $R$. The solid line is an \emph{ab initio} calculation, without adjustable parameter (the shaded area arises from uncertainty in the calibration of $R$). Figure adapted from~\cite{Barredo2015}.}
\label{fig:XY}
\end{figure}

The pair of atoms thus prepared in $\ket{pd}$ is left to evolve for an adjustable time $T$ before the state of the system is readout by sending a Rydberg pulse which de-excites the $\ket{d}$ state back to the ground state, while $\ket{p}$ remains unaffected. Figure~\ref{fig:XY}a shows coherent oscillations of the populations of the $\ket{pd}$ and $\ket{dp}$ states as a function of $T$. The oscillation frequency varies as $1/R^3$ (Figure~\ref{fig:XY}b), as expected for this dipolar-induced excitation exchange. 

\subsection{Conclusion on the measurement of interactions between Rydberg states}

\begin{figure}[t]
\centering
\includegraphics[width=5.8cm]{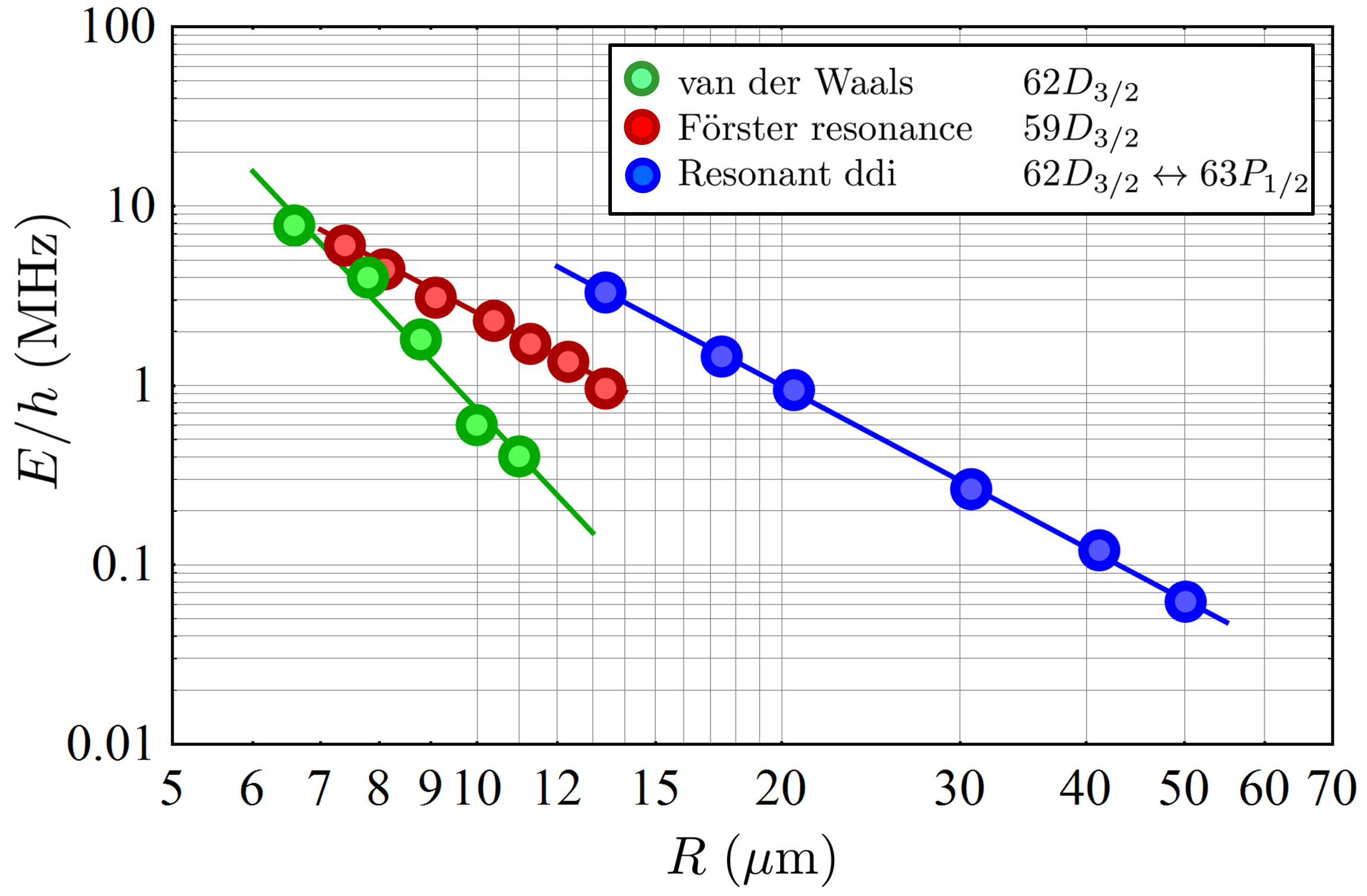}
\caption{Summary of the measurements of the interactions between Rydberg atoms versus distance, in various regimes, performed at Institut d'Optique. The atoms are separated by a distance $R$, the internuclear axis being aligned with the quantization axis. Disks are the measured values, and solid lines the theoretical interaction without any adjustable parameter. }
\label{fig:summary_exp}
\end{figure}

\begin{figure}[t]
\centering
\includegraphics[width=5cm]{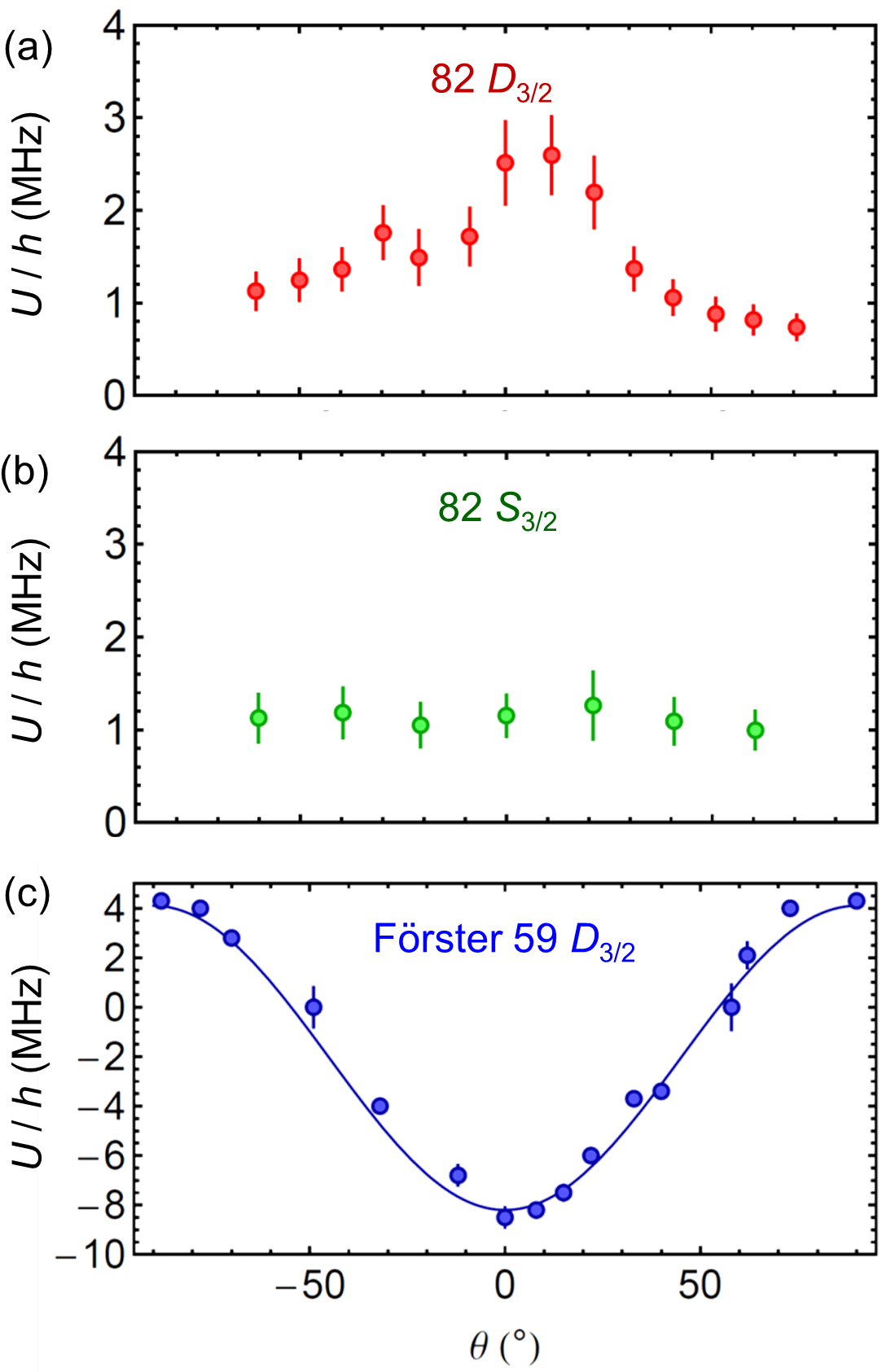}
\caption{Angular dependence of the interactions between two Rydberg atoms. The internuclear axis and the quantization axis are at an angle $\theta$. The van der Waals interaction between two atoms in $\ket{nD_{3/2}}$ shows a significant anisotropy (a), while it is isotropic for $\ket{nS_{1/2}}$ states (b). (c) At a F\"orster resonance, the interaction shows the characteristic angular pattern $\propto\left(1-3\cos^2\theta\right)$ of the dipole-dipole interaction (solid line). Figure adapted from~\cite{Barredo2014,Ravets2015}.}
\label{fig:angle}
\end{figure}

This series of experiments allowed to explore in detail the spatial dependence  of the various types of interactions between Rydberg atoms, both as a function of distance (Fig.~\ref{fig:summary_exp}), and as a function of the angle (Fig.~\ref{fig:angle}). The very good agreement between theory experiments  shows that the experimental control of small systems of single atoms excited to Rydberg states is good enough for such studies to be extended to larger number of atoms, as we shall describe in the next section.

\section{Towards larger number of atoms}\label{Sec:large_atom_numbers}

The direct measurement and control of the interactions between Rydberg atoms in electric and magnetic fields shown above enables the quantum simulation of complex synthetic quantum systems in arbitrary geometries. Indeed, besides the demonstration of a CNOT gate between two atoms in an array of micro traps mentioned in Section~\ref{Sec:exp_entanglement_CNOT}~\cite{Maller2015}, two groups recently performed experiments where more than 2 atoms interact with each other. Both engineer and simulate spin Hamiltonians, as described in Section~\ref{Sec:theo_interaction}. Beyond proof-of-principle demonstrations, these experiments allowed capturing the main technical imperfections and quantify their effects on the spin dynamics. 

\subsection{Ising dynamics in three-atom systems~\cite{Barredo2014}}

\begin{figure}[t]
\centering
\includegraphics[width=8.5cm]{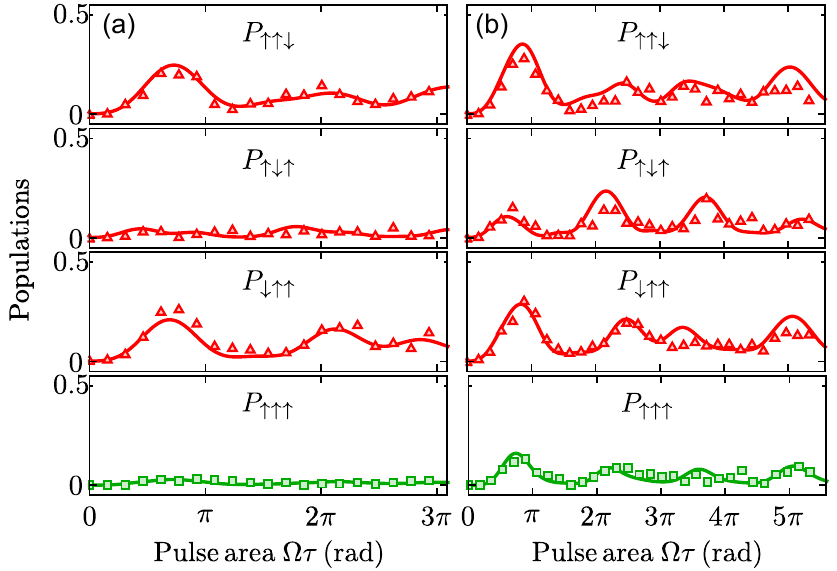}
\caption{Simulating a quantum Ising magnet with three Rydberg atoms. (a): Probability distributions $P_{{\uparrow\uparrow\downarrow}}$, $P_{{\uparrow\downarrow\uparrow}}$,$P_{{\downarrow\uparrow\uparrow}}$, and $P_{{\uparrow\uparrow\uparrow}}$ averaged over 100 realizations of the experiment for a Rabi frequency $\Omega = 2 \pi \times 0.8$ MHz. The atoms are separated by 12 $\mu$m and arranged in an equilateral triangle with one side aligned with the quantization axis $\hat{z}$. (b): Populations for $\Omega = 2 \pi \times 1.6$ MHz. Figure adapted from~\cite{Barredo2014}.}
\label{fig:Ising_3atoms}
\end{figure}

As a first example, the Institut d'Optique group implemented the Ising-like Hamiltonian~(\ref{eq:ising}) for a system of three spins arranged in an equilateral triangle~\cite{Barredo2014}. To highlight the opportunities anisotropic interactions might bring, the group excited the atoms to the $\ket{82 D_{3/2},m_j=3/2}$ Rydberg state. For atoms separated by 12 $\mu$m and a driving Rabi frequency of $2 \pi \times 0.8$ MHz, the van der Waals blockade is only partial, as  the atom pairs exhibit effective interaction energies $(V_{12}, V_{23}, V_{13})= h \times (0.9, 1.1, 2.6)$ MHz. The experiment started by initializing the system to the state $\ket{\!\!\downarrow\downarrow\downarrow}$. Then, applying the excitation laser for a variable time, the final spin state was measured. The result is shown in Fig. \ref{fig:Ising_3atoms}b, where the angular dependence of $V_{\rm eff}$ becomes apparent in the dynamics.

The probability to excite the state $\ket{\!\!\uparrow \downarrow\uparrow}$ is almost totally suppressed, while it is appreciable for both $\ket{\!\!\uparrow\uparrow\downarrow}$ and $\ket{\!\!\downarrow\uparrow\uparrow}$, which show very similar dynamics. Increasing the Rabi frequency to $2 \pi \times 1.6$ MHz partially overcomes the blockade of triple excitations, but the asymmetry in the curves for double excitations due to anisotropic interactions can still be observed. Solid lines represent the predicted dynamics of the two level system evolving under the Hamiltonian~(\ref{eq:ising}), with no adjustable parameters. The simulation includes the independently measured Rabi frequencies and damping rates for single atom in each site. The small damping rates observed are mainly due to off-resonant spontaneous emission through the intermediate state $\ket{5P_{1/2}}$. In addition, the numerical results account for the effect of $\sim5\%$ atom losses in the populations. The agreement is very good and demonstrates the promises of cold Rydberg atoms to perform quantum simulations of Ising Hamiltonians.

\subsection{XY Hamiltonian dynamics in chain of three-atoms~\cite{Barredo2015}}

In a second experiment, the Institut d'Optique group used resonant dipole-dipole interactions to engineer the XY Hamiltonian~(\ref{eq:XY_hamiltonian}) for a chain of three Rydberg atoms aligned along the quantization axis~\cite{Barredo2015}. In this configuration, two different pairwise interaction strengths are at play. Owing to the $1/R^3$ scaling of the dipole-dipole interaction, the coupling is 8 times as large for nearest neighbors as for the two furthermost atoms. As a consequence, when the system is initialized in the state $\ket{\!\!\uparrow\downarrow\downarrow}$, the eigenvalues of the Hamiltonian are incommensurate and the dynamics is expected to show aperiodic oscillations in the populations of $\ket{\!\!\uparrow\downarrow\downarrow}$ and $\ket{\!\!\downarrow\downarrow\uparrow}$. 

\begin{figure}[t]
\centering
\includegraphics[width=8cm]{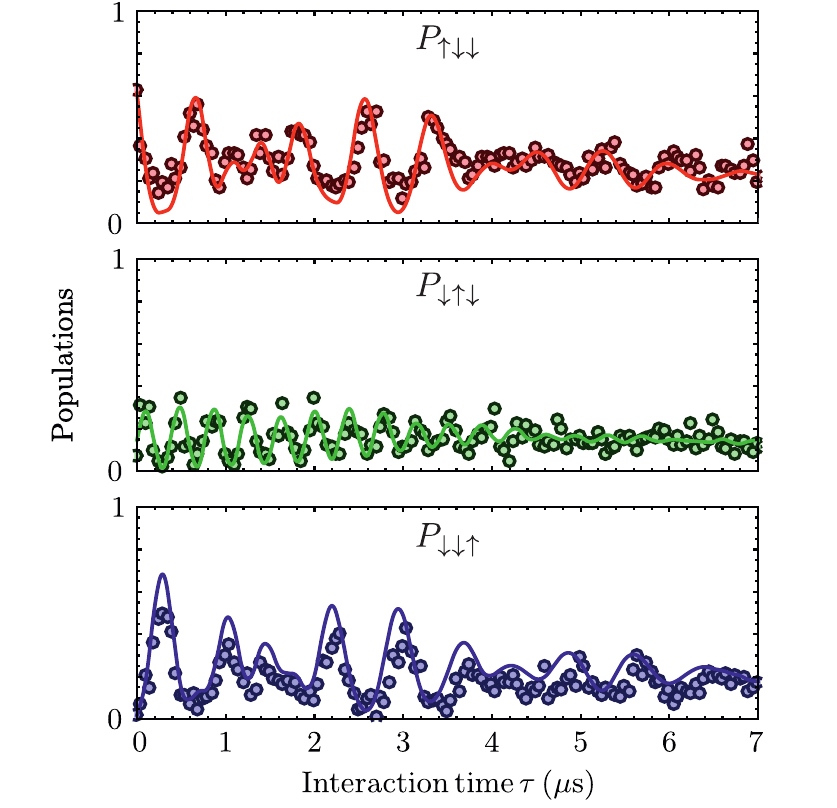}
\caption{Coherent excitation hopping in a spin chain. Dynamics for a system initially prepared in the state $\ket{\!\!\uparrow\downarrow\downarrow}$ and evolving under the Hamiltonian  (\ref{eq:XY_hamiltonian}). The atom are separated by 20 $\mu$m and aligned with the quantization axis. Disks are experimental data points averaged over 100 realizations. Curves represent the predicted dynamics taking into account experimental imperfections, without any adjustable parameter. Figure adapted from~\cite{Barredo2015}.}
\label{fig:spin_chain}
\end{figure}

This is qualitatively observed in the experimental data shown in Fig.~\ref{fig:spin_chain}b, which exhibits collapse and revivals in the dynamics due to the long-range coupling. There, solid lines are the result of a numerical simulation of the XY Hamiltonian (\ref{eq:XY_hamiltonian}), including finite preparation fidelities, atom temperature effects, and detection errors. Here also, the agreement with the experimental data is very good and shows that the system can be effectively reduced to a three-particle two-level model. Moreover, since there are not fundamental limitations in sight to reduce the effect of imperfections, this result strengthens the ambition to perform larger scale quantum simulations with Rydberg atoms.

\begin{figure}[t]
\centering
\includegraphics[width=8.5cm]{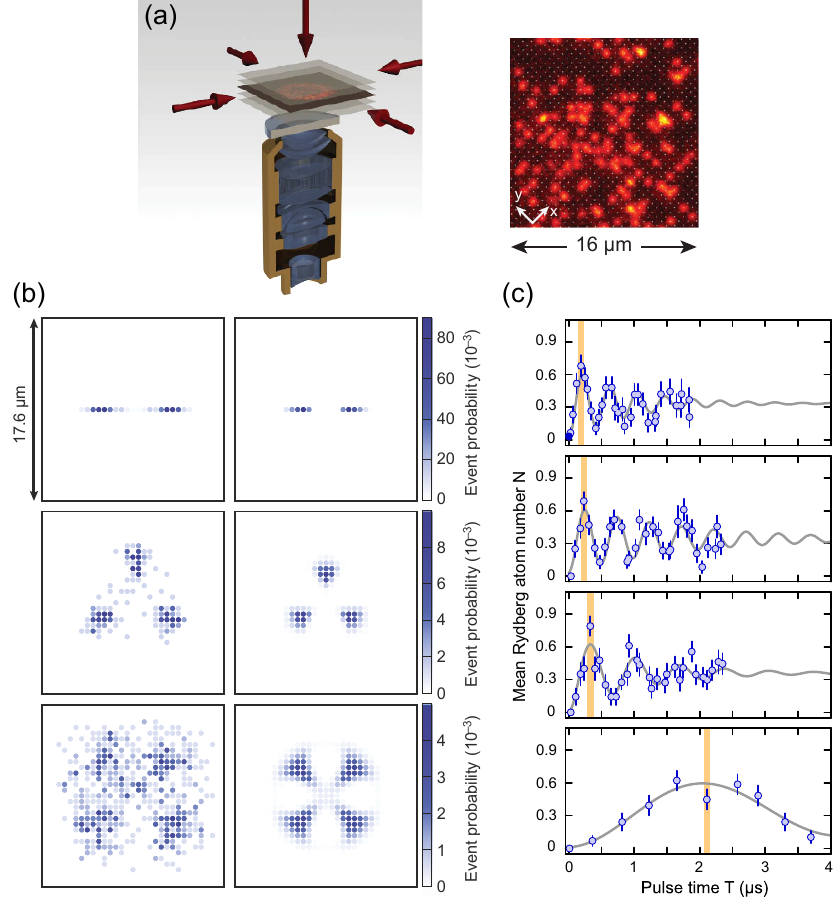}
\caption{Rydberg atoms in optical lattices. (a): A high-resolution microscope objective allows the observation of individual atoms in 2D. (b): The compilation of many single shot images of Rydberg states results in spatially ordered structures (left), in good agreement with the theoretical prediction (right). (c): Observed $\sqrt{N}$ enhancement of the coupling of the atoms with the light field. Figure adapted from~\cite{Sherson2010,Schauss2012,Zeiher2015}.}
\label{fig:Bloch}
\end{figure}

\subsection{Blockade in optical lattices} 

Although not directly the focus of this review, we briefly mention a  series of  experiments involving Rydberg atoms prepared in 2D optical lattices, that have been performed in Munich since 2012. In these experiments, which also implement the Ising-like Hamiltonian~(\ref{eq:ising}), single atoms are trapped in 2D arrays of $\sim 10\times 10$ sites created by optical potentials, and imaged using a high resolution microscope objective, as depicted in Fig.~\ref{fig:Bloch}a. In this experiment, the distance between the atoms is $a=500$ nm. 

By shining the excitation laser on all the atoms at the same time, the group demonstrated the Rydberg blockade in their system by observing spatially ordered structures (Fig.~\ref{fig:Bloch}b)~\cite{Schauss2012}. For this particular demonstration, the fact that the atoms are arranged in a two-dimensional periodic structure is irrelevant, as each blockade sphere contains many atoms ($R_{\rm b}\sim 10 a$). However, the high resolution microscope provides a spatial detection of the Rydberg excitation, which is challenging to achieve in cold atomic ensemble with random positions of the atoms~\cite{Gunter2012}.  

The group also coherently manipulated a collective state composed of up to 185 individual atoms, and confirmed the expected $\sqrt{N}$ enhancement of the Rabi frequency (Fig.~\ref{fig:Bloch}c)~\cite{Zeiher2015}. Recently, the same group succeeded in the preparation of a state closed to a Rydberg crystal with a precise number of excitations via adiabatic sweeps of the laser parameters~\cite{Schauss2015}. 

\section{Conclusion}

Systems of individually trapped and addressed Rydberg atoms enter an exciting time. After the recent demonstration of elementary building blocks, they should provide in the coming years an ideal platform to study many-body physics in the laboratory, with many possible applications in quantum simulation and quantum information processing. 

\section*{Acknowledgments}
We thank L. B\'eguin, A. Vernier, S. Ravets, H. Labuhn,  S. de L\'es\'eleuc, F. Nogrette, A. Ga\"etan, C. Evellin, T. Wilk, Y. Miroshnychenko, R. Chicireanu, P. Grangier, P. Pillet, D. Comparat, C.S. Adams, Y.R.P. Sortais, and J. Wolters for their contributions to the experiments reported in this review. We acknowledge funding by the EU [ERC Stg Grant ARENA, FET-Open Xtrack Project HAIRS, H2020 FET-PROACT Project RySQ, and EU Marie-Curie Program ITN COHERENCE], by the `PALM' Labex (project QUANTICA) and by the Region \^Ile-de-France in the framework of DIM Nano-K and IFRAF. We thank Mark Saffman and Grant Biedermann for careful reading of the manuscript.


\section*{References}

\begin{thebibliography}{100}

\bibitem{Gallagher1994}
T.F. Gallagher,
\emph{Rydberg atoms}, 
Cambridge University Press (Cambridge, UK, 1994). 

\bibitem{Jaksch2000}  
D. Jaksch, J.I. Cirac, P. Zoller, S.L. Rolston, R. C\^ot\'e, and M.D. Lukin,
\emph{Fast quantum gates for neutral atoms}, 
\doilink{10.1103/PhysRevLett.85.2208}{Phys. Rev. Lett {\bf 85}, 2208 (2000)}.

\bibitem{Lukin2001} 
M.D. Lukin, M. Fleischhauer, R.~Cot\'{e}, L.M.~Duan, D.~Jaksch, J.I.~Cirac and P.~Zoller, 
\emph{Dipole Blockade and Quantum Information Processing in Mesoscopic Atomic Ensembles},
\doilink{10.1103/PhysRevLett.87.037901}{Phys. Rev. Lett. {\bf 87}, 037901 (2001)}.

\bibitem{Saffman2010} 
M.~Saffman, T.G.~Walker and K.~M{\o}lmer, 
\emph{Quantum information with Rydberg atoms},
\doilink{10.1103/RevModPhys.82.2313}{Rev. Mod. Phys. {\bf 82}, 2313 (2010)}.

\bibitem{Pritchard2013}
J.D. Pritchard, K.J. Weatherill and C.S. Adams, 
\emph{Non-linear optics using cold Rydberg atoms}
\doilink{10.1142/9789814440400_0008}{Annual Review of Cold Atoms and Molecules, {\bf 1}, 301 (2013)}.

\bibitem{Comparat2010}
D. Comparat and P. Pillet, 
\emph{Dipole blockade in a cold Rydberg atomic sample},
\doilink{10.1364/JOSAB.27.00A208}{J. Opt. Soc. Am. B {\bf 27}, A208 (2010)}.

\bibitem{Schlosser2001} 
N. Schlosser, G. Reymond, I. Protsenko and P. Grangier, 
\emph{Sub-poissonian loading of single atoms in a microscopic dipole trap}, 
\doilink{10.1038/35082512}{Nature {\bf 411}, 1024 (2001)}.

\bibitem{Schauss2012} 
P. Schau\ss, M. Cheneau, M. Endres, T. Fukuhara, S. Hild, A. Omran, T. Pohl, C. Gross, S. Kuhr, and I. Bloch,
\emph{Observation of spatially ordered structures in a two-dimensional Rydberg gas},
\doilink{10.1038/nature11596}{Nature {\bf 491}, 87 (2012)}.

\bibitem{Loew2013}
R. L\"ow, H. Weimer, J. Nipper, J. B. Balewski, B. Butscher, H. P. Büchler, and T. Pfau,
\emph{An experimental and theoretical guide to strongly interacting Rydberg gases},
\doilink{10.1088/0953-4075/45/11/113001}{J. Phys. B {\bf 45}, 113001 (2012)}.

\bibitem{Raimond1981}
J.M. Raimond, G. Vitrant and S. Haroche,
\emph{Spectral line broadening due to the interaction between very excited atoms: 'the dense Rydberg gas'}
\doilink{10.1088/0022-3700/14/21/003}{J. Phys. B: At. Mol. Phys. {\bf 14} L655 (1981)}.

\bibitem{Haroche2013}
S. Haroche,
\emph{Nobel Lecture: Controlling photons in a box and exploring the quantum to classical boundary}
\doilink{10.1103/RevModPhys.85.1083}{Rev. Mod. Phys. {\bf 85}, 1083 (2013)}.

\bibitem{Anderson1998} 
W.R. Anderson, J.R. Veale, and T.F. Gallagher,
\emph{Resonant dipole-dipole energy transfer in a nearly frozen Rydberg gas},
\doilink{10.1103/PhysRevLett.80.249}{Phys. Rev. Lett. {\bf 80}, 249 (1998)}.

\bibitem{Mourachko1998} 
I. Mourachko, D. Comparat, F. de Tomasi, A. Fioretti, P. Nosbaum,  V. Akulin, and P. Pillet,
\emph{Many-body effects in a frozen Rydberg gas},
\doilink{10.1103/PhysRevLett.80.253}{Phys. Rev. Lett. {\bf 80}, 253 (1998)}.

\bibitem{Saffman2005}
M. Saffman and T. G. Walker,
\emph{Analysis of a quantum logic device based on dipole-dipole interactions of optically trapped Rydberg atoms},
\doilink{10.1103/PhysRevA.72.022347}{Phys. Rev. A {\bf 72}, 022347 (2005)}.

\bibitem{Mandel2003}
O. Mandel, M. Greiner, A. Widera, T. Rom, T.W. H\"ansch and I. Bloch,
\emph{Controlled collisions for multi-particle entanglement of optically trapped atoms}, 
\doilink{10.1038/nature02008}{Nature \textbf{425}, 937 (2003)}.

\bibitem{Saffman2002} 
M.~Saffman and T.G.~Walker, 
\emph{Creating single-atom and single-photon sources from entangled atomic ensembles},
\doilink{10.1103/PhysRevA.66.065403}{Phys. Rev. A {\bf 66}, 065403 (2002)}.

\bibitem{Muller2014} 
M.M. M\"uller, M. Murphy, S. Montangero, T. Calarco, P. Grangier and A. Browaeys,
\emph{Implementation of an experimentally feasible controlled-phase gate
on two blockaded Rydberg atoms},
\doilink{10.1103/PhysRevA.89.032334}{Phys. Rev. A {\bf 89}, 032334 (2014)}.

\bibitem{Saffman2009}
M. Saffman and K. Moelmer, 
\emph{Efficient multiparticle entanglement via asymmetric Rydberg blockade}, 
\doilink{10.1103/PhysRevLett.102.240502}{Phys. Rev. Lett. {\bf 102}, 240502 (2009)}.

\bibitem{Isenhower2011}
L. Isenhower, M. Saffman, and K. Mølmer, 
\emph{Multibit CkNOT quantum gates via Rydberg blockade}, 
\doilink{10.1007/s11128-011-0292-4}{Quant. Inf. Proc. {\bf 10}, 755 (2011)}.

\bibitem{Mueller2009} 
M. M\"uller, I. Lesanovsky, H. Weimer, H. P. B\"uchler, and P. Zoller,
\emph{Mesoscopic Rydberg Gate Based on Electromagnetically Induced Transparency},
\doilink{10.1103/PhysRevLett.102.170502}{Phys. Rev. Lett. \textbf{102}, 170502 (2009)}.

\bibitem{Moller2008} 
D. Moeller, L.B. Madsen, and K. Moelmer, 
\emph{Quantum Gates and Multiparticle Entanglement by Rydberg Excitation Blockade and Adiabatic Passage},
\doilink{10.1103/PhysRevLett.100.170504}{Phys. Rev. Lett. \textbf{100}, 170504 (2008)}.

\bibitem{Nielsen2000}
M.A. Nielsen and I.L. Chuang, 
\emph{Quantum computation and quantum information},
Cambridge University Press (Cambridge, 2000).

\bibitem{Feynman1982}
R. Feynman, 
\emph{Simulating Physics with computers},
\href{http://link.springer.com/article/10.1007%2FBF02650179#page-1}{Int. J. Theor. Phys. {\bf 21}, 467-488 (1982)}.

\bibitem{Lloyd1996}
S. Lloyd,
\emph{Universal Quantum Simulators},
\doilink{10.1126/science.273.5278.1073}{Science 273, 1073(1996)}.

\bibitem{Georgescu2014}
I.M. Georgescu, S. Ashhab, F. Nori, 
\emph{Quantum simulation}, 
\doilink{10.1103/RevModPhys.86.153}{Rev. Mod. Phys. {\bf 86}, 153 (2014)}.

\bibitem{Weimer2010} 
H. Weimer, M. M\"uller, I. Lesanovsky, P. Zoller and H.P. B\"uchler, 
\emph{A Rydberg quantum simulator},
\doilink{10.1038/nphys1614}{Nat. Phys. {\bf 6}, 382 (2010)}.

\bibitem{Reinhard2007} 
A. Reinhard, T. Cubel Liebisch, B. Knuffman, and G. Raithel,
\emph{Level shifts of rubidium Rydberg states due to binary interactions},
\doilink{10.1103/PhysRevA.75.032712}{Phys. Rev. {\bf 75}, 032712 (2007)}.

\bibitem{Cano2012}
D. Cano and J. Fort\'agh,
\emph{Nonadditive potentials between three Rydberg atoms},
\doilink{10.1103/PhysRevA.86.043422}{Phys. Rev. A {\bf 86}, 043422 (2012)}.

\bibitem{Walker2008}
T.G. Walker and M. Saffman, 
\emph{Consequences of Zeeman degeneracy for the van der Waals blockade between Rydberg atoms},
\doilink{10.1103/PhysRevA.77.032723}{Phys. Rev. A {\bf 77}, 032723 (2008)}.

\bibitem{Vermersch2015a}
B. Vermersch, A.W. Glaetzle, and P. Zoller,
\emph{Magic distances in the blockade mechanism of Rydberg P and D states},
\doilink{10.1103/PhysRevA.91.023411}{Phys. Rev. A {\bf 91}, 023411 (2015)}.

\bibitem{Saffman2005b}
T.G. Walker and M. Saffman, 
\emph{Zeros of Rydberg-Rydberg F\"oster interactions},
\doilink{10.1088/0953-4075/38/2/022}{J. Phys. B {\bf 38}, S309 (2005)}. 

\bibitem{Beterov2015}
I. I. Beterov and M. Saffman, 
\emph{Rydberg blockade, Förster resonances, and quantum state measurements with different atomic species}, 
\doilink{10.1103/PhysRevA.92.042710}{Phys. Rev. A {\bf 92}, 042710 ( 2015)}.

\bibitem{Forster1948}
T. F\"orster,
\emph{Zwischenmolekulare Energiewanderung und Fluoreszen}, 
\doilink{10.1002/andp.19484370105}{Ann. Phys. {\bf 437}, 55 (1948).}

\bibitem{Clegg2006}
R.M. Clegg,
\emph{The History of Fret: From conception through the labors of birth}
\doilink{10.1007/0-387-33016-X_1}{\emph{in} Reviews in Fluorescence 2006, 1 (Springer, 2006)}.

\bibitem{Nipper2012a}
J. Nipper, J.B. Balewski, A.T. Krupp, B. Butscher, R. L\"ow, and T. Pfau
\emph{Highly Resolved Measurements of Stark-Tuned F\"orster Resonances between Rydberg Atoms},
\doilink{10.1103/PhysRevLett.108.113001}{Phys. Rev. Lett. {\bf 108}, 113001 (2012)}.

\bibitem{Nipper2012b}
J. Nipper, J. B. Balewski, A. T. Krupp, S. Hofferberth, R. L\"ow, and T. Pfau
\emph{Atomic Pair-State Interferometer: Controlling and Measuring an Interaction-Induced Phase Shift in Rydberg-Atom Pairs},
\doilink{10.1103/PhysRevX.2.031011}{Phys. Rev. X {\bf 2}, 031011 (2012)}.

\bibitem{Ravets2015}
S. Ravets, H. Labuhn, D. Barredo, T. Lahaye, and A. Browaeys,
\emph{Measurement of the angular dependence of the dipole-dipole interaction between two individual Rydberg atoms at a F\"orster resonance},
\doilink{10.1103/PhysRevA.92.020701}{Phys. Rev. A {\bf 92}, 020701(R) (2015)}.

\bibitem{Hauke2010}
P. Hauke, F.M. Cucchietti, A. M\"uller-Hermes, M.-C. Ba\~nuls, J.I. Cirac, and M. Lewenstein, 
\emph{Complete devil's staircase and crystal-superfluid transitions in a dipolar XXZ spin chain: a trapped ion quantum simulation},
\doilink{10.1088/1367-2630/12/11/113037}{New J. Phys. {\bf 12}, 113037 (2010)}.

\bibitem{Peter2012}
D. Peter, S. M\"uller, S. Wessel, and H. P. B\"uchler, 
\emph{Anomalous Behavior of Spin Systems with Dipolar Interactions},
\doilink{10.1103/PhysRevLett.109.025303}{Phys. Rev. Lett. {\bf 109}, 025303 (2012)}.

\bibitem{Yan2013}
B. Yan, S.A. Moses, B. Gadway, J.P. Covey, K.R.A. Hazzard, A.M. Rey, D.S. Jin, and J. Ye, 
\emph{Observation of dipolar spin-exchange interactions with lattice-confined polar molecules},
\doilink{10.1038/nature12483}{Nature {\bf 501}, 521 (2013)}.

\bibitem{Paz2013}
A. de Paz, A. Sharma, A. Chotia, E. Mar\'echal, J. H. Huckans, P. Pedri, L. Santos, O. Gorceix, L. Vernac, and B. Laburthe-Tolra,
\emph{Nonequilibrium Quantum Magnetism in a Dipolar Lattice Gas},
\doilink{10.1103/PhysRevLett.111.185305}{Phys. Rev. Lett. {\bf 111}, 185305 (2013)}.

\bibitem{Zimmerman1979}
M.L. Zimmerman, M.G. Littman, M.M. Kash, and D. Kleppner,
\emph{Stark structure of the Rydberg states of alkali-metal atoms},
\doilink{10.1103/PhysRevA.20.2251}{Phys. Rev. A {\bf 20}, 2251 (1979)}.

\bibitem{Derevianko2015}
A. Derevianko, P. K\'om\'ar, T. Topcu, R.M. Kroeze, and M.D. Lukin,
\emph{Effects of molecular resonances on Rydberg blockade},
\doilink{10.1103/PhysRevA.92.063419}{Phys. Rev. A {\bf 92}, 063419 (2015)}.

\bibitem{Hankin2014}
A.M. Hankin, Y.-Y. Jau, L.P. Parazzoli, C.W. Chou, D.J. Armstrong, A.J. Landahl, and G.W. Biedermann,
\emph{Two-atom Rydberg blockade using direct $6S$ to $nP$ excitation}, 
\doilink{10.1103/PhysRevA.89.033416}{Phys. Rev. A {\bf 89}, 033416 (2014)}.

\bibitem{Grimm1998}
R. Grimm, M. Weidem\"uller, and Yu.B. Ovchinnikov,
\emph{Optical Dipole Traps for Neutral Atoms}, 
\doilink{10.1016/S1049-250X(08)60186-X}{Adv. At. Mol. Opt. Phys. {\bf 42} 95 (2000)}.

\bibitem{Hu1994}
Z. Hu and H.J. Kimble, 
\emph{Observation of a single atom in a magneto-optical trap},
\doilink{10.1364/OL.19.001888}{Opt. Lett. {\bf 19}, 1888 (1994)}.

\bibitem{Kuhr2001}
S. Kuhr, W. Alt, D. Schrader, M. M\"uller, V. Gomer, D. Meschede,
\emph{Deterministic Delivery of a Single Atom},
\doilink{10.1126/science.1062725}{Science {\bf 293}, 278 (2001)}. 

\bibitem{Meschede2006}
D. Meschede and A. Rauschenbeutel, 
\emph{Manipulating Single Atoms}
\doilink{10.1016/S1049-250X(06)53003-4}{ Adv. At. Mol. Opt. Phys. {\bf 53}, 75 (2006)}.

\bibitem{Alt2002}
W. Alt, 
\emph{An objective lens for efficient fluorescence detection of single atoms},
\doilink{10.1078/0030-4026-00133}{Optik {\bf 113}, 142 (2002)}.

\bibitem{Sortais2007} 
Y.R.P.~Sortais, H.~Marion, C.~Tuchendler, A.M.~Lance, M.~Lamare, P.~Fournet, C.~Armellin, R.~Mercier, G.~Messin, A.~Browaeys and P.~Grangier,
\emph{Diffraction-limited optics for single-atom manipulation}, 
\doilink{10.1103/PhysRevA.75.013406}{Phys. Rev. A \textbf{75}, 013406 (2007)}.

\bibitem{Fuhrmanek2012} 
A.~Fuhrmanek, R.~Bourgain, Y.R.P.~Sortais and A.~Browaeys,
\emph{Light-assisted collisions between a few cold atoms in a microscopic dipole trap},
\doilink{10.1103/PhysRevA.85.062708}{Phys. Rev. A \textbf{85}, 062708 (2012)}.

\bibitem{Schlosser2002} 
N.~Schlosser, G.~Reymond and P.~Grangier, 
\emph{Collisional blockade in microscopic optical dipole traps},
\doilink{10.1103/PhysRevLett.89.023005}{Phys. Rev. Lett. {\bf 89}, 023005 (2002)}.

\bibitem{Ebert2014}
M. Ebert, A. Gill, M. Gibbons, X. Zhang, M. Saffman, and T.G. Walker,
\emph{Atomic Fock State Preparation Using Rydberg Blockade},
\doilink{10.1103/PhysRevLett.112.043602}{Phys. Rev. Lett. \textbf{112}, 043602 (2014)}.

\bibitem{Grunzweig2010}
T. Gr\"unzweig, A. Hilliard, M. McGovern, and M.F. Andersen, 
\emph{Near-deterministic preparation of a single atom in an optical microtrap}
\doilink{10.1038/NPHYS1778}{Nat. Phys. {\bf 6}, 951 (2010)}.

\bibitem{Carpentier2013}
A.V.  Carpentier, Y.H. Fung, P. Sompet, A.J. Hilliard, T.G. Walker, and M.F. Andersen,
\emph{Preparation of a single atom in an optical microtrap},
\doilink{10.1088/1612-2011/10/12/125501} {Las. Phys. Lett. {\bf 10}, 125501 (2013)}.

\bibitem{Fung2015}
Y.H. Fung, Y. H. and M.F. Andersen, 
\emph{Efficient collisional blockade loading of a single atom into a tight microtrap}, 
\doilink{10.1088/1367-2630/17/7/073011}{New J. Phys. {\bf 17}, 073011 (2015)}.

\bibitem{Lester2015}
B.J. Lester, N. Luick, A.M. Kaufman, C.M. Reynolds, and C.A. Regal,
\emph{Rapid Production of Uniformly Filled Arrays of Neutral Atoms},
\doilink{10.1103/PhysRevLett.115.073003}{Phys. Rev. Lett. \textbf{115}, 073003 (2015)}.

\bibitem{Nelson2007}
K.D. Nelson, X. Li, and D.S. Weiss,
\emph{Imaging single atoms in a three-dimensional array},
\doilink{10.1038/nphys645}{Nat. Phys. {\bf 3}, 556 (2007)}.

\bibitem{Wang2015}
Y. Wang, X. Zhang, T.A. Corcovilos, A. Kumar, and D.S. Weiss,
\emph{Coherent Addressing of Individual Neutral Atoms in a 3D Optical Lattice},
\doilink{10.1103/PhysRevLett.115.043003}{Phys. Rev. Lett. {\bf 115}, 043003 (2015)}.

\bibitem{Bakr2009}
W.S. Bakr, J.I. Gillen, A. Peng, S. Foelling, and M. Greiner,
\emph{A Quantum Gas Microscope for detecting single atoms in a Hubbard regime optical lattice},
\doilink{10.1038/nature08482}{Nature {\bf 462}, 74 (2009)}.

\bibitem{Sherson2010}
J.F. Sherson, C. Weitenberg, M. Endres, M. Cheneau, I. Bloch, and S. Kuhr,
\emph{Single-atom-resolved fluorescence imaging of an atomic Mott insulator},
\doilink{10.1038/nature09378}{Nature {\bf 467}, 68 (2010)}.

\bibitem{Lee2007}
P.J. Lee, M. Anderlini, B.L. Brown, J. Sebby-Strabley, W.D. Phillips, J.V. Porto,
\emph{Sublattice addressing and spin-dependent motion of atoms in a double-well lattice}
\doilink{10.1103/PhysRevLett.99.020402}{Phys. Rev. Lett. {\bf 99}, 020402 (2007)}.

\bibitem{Lee2013}
J.H. Lee, E. Montano, I.H. Deutsch, and P.S. Jessen,
\emph{Robust site-resolvable quantum gates in an optical lattice via inhomogeneous control},
\doilink{10.1038/ncomms3027}{Nat. Comm. {\bf 4}, 2027 (2013)}.

\bibitem{Weitenberg2011}
C. Weitenberg, M. Endres, J. F. Sherson, M. Cheneau, P. Schau{\ss}, T. Fukuhara, I. Bloch, and S. Kuhr,
\emph{Single-spin addressing in an atomic Mott insulator},
\doilink{10.1038/nature09827}{Nature {\bf 471}, 319 (2011)}.

\bibitem{Bergamini2004}
S. Bergamini, B. Darqui\'{e}, M. Jones, L. Jacubowiez, A. Browaeys, and P. Grangier,
\emph{Holographic generation of microtrap array for single atoms by use of a programmable phase modulator},
\doilink{10.1364/JOSAB.21.001889}{J. Opt. Soc. Am. B {\bf 21}, 001889 (2004)}.

\bibitem{Nogrette2013}
F. Nogrette, H. Labuhn, S. Ravets, D. Barredo, L. Béguin, A. Vernier, T. Lahaye, and A. Browaeys,
\emph{Single-atom trapping in holographic arrays of micro traps with arbitrary geometries}, 
\doilink{10.1103/PhysRevX.4.021034}{Phys. Rev. X {\bf 4}, 021034 (2014)}.

\bibitem{Knoernschild2010}
C. Knoernschild, X. L. Zhang, L. Isenhower, A. T. Gill, F. P. Lu, M. Saffman, and J. Kim,
\emph{Independent individual addressing of multiple neutral atom qubits with a micromirror-based beam steering system},
\doilink{10.1063/1.3494526}{Appl. Phys. Lett. {\bf 97}, 134101 (2010)}.

\bibitem{Dumke2002} 
R.~Dumke, M.~Volk, T.~M\"uther, F.B.J.~Buchkremer, G.~Birkl and W.~Ertmer, 
\emph{Micro-optical Realization of Arrays of Selectively Addressable Dipole Traps: A Scalable Configuration for Quantum Computation with Atomic Qubits},
\doilink{10.1103/PhysRevLett.89.097903}{Phys. Rev. Lett. {\bf 89}, 097903 (2002)}.

\bibitem{Labuhn2014} 
H. Labuhn, S. Ravets, D. Barredo, L. B\'eguin, F. Nogrette, T. Lahaye, and A. Browaeys,
\emph{Single-atom addressing in microtraps for quantum-state engineering using Rydberg atoms}
\doilink{10.1103/PhysRevA.90.023415}{Phys. Rev. A {\bf 90}, 023415 (2014)}.

\bibitem{Schlosser2011}
M. Schlosser, S. Tichelmann, J. Kruse, G. Birkl,
\emph{Scalable architecture for quantum information processing with atoms in optical micro-structures},
\doilink{10.1007/s11128-011-0297-z}{Quantum Inf Process {\bf 10}, 907 (2011)}.

\bibitem {Piotrowicz2013}
M.J. Piotrowicz, M. Lichtman, K. Maller, G. Li, S. Zhang, L. Isenhower, and M. Saffman,
\emph{Two-dimensional lattice of blue-detuned atom traps using a projected Gaussian beam array},
\doilink{10.1103/PhysRevA.88.013420}{Phys. Rev. A {\bf 88}, 013420 (2013)}.

\bibitem {Zhang2011}
S. Zhang, F. Robicheaux, and M. Saffman,
\emph{Magic-wavelength optical traps for Rydberg atoms},
\doilink{10.1103/PhysRevA.84.043408}{Phys. Rev. A {\bf 84}, 043408 (2011)}.

\bibitem{Johnson2008}
T. A. Johnson, E. Urban, T. Henage, L. Isenhower, D. D. Yavuz, T. G. Walker, and M. Saffman,
\emph{Rabi Oscillations between Ground and Rydberg States with Dipole-Dipole Atomic Interactions},
\doilink{10.1103/PhysRevLett.100.113003}{Phys. Rev. Lett. {\bf 100}, 113003 (2008)}.

\bibitem{miroshnychenko2010}
Y. Miroshnychenko, A. Ga\"etan, C. Evellin, P. Grangier, D. Comparat, P. Pillet, T. Wilk, and A. Browaeys,
\emph{Coherent excitation of a single atom to a Rydberg state},
\doilink{10.1103/PhysRevA.82.013405}{Phys. Rev. A {\bf 82}, 013405 (2010)}.

\bibitem{Zuo2009}
Zhanchun Zuo, Miho Fukusen, Yoshihito Tamaki, Tomoki Watanabe, Yusuke Nakagawa and Ken'ichi Nakagawa,
\emph{Single atom Rydberg excitation in a small dipole trap},
\doilink{10.1364/OE.17.022898}{Optics Express {\bf 25}, 22898 (2009)}.

\bibitem{Maller2015}
K.M. Maller, M.T. Lichtman, T. Xia, Y. Sun, M.J. Piotrowicz, A.W. Carr, L. Isenhower, and M. Saffman,
\emph{Rydberg-blockade controlled-not gate and entanglement in a two-dimensional array of neutral-atom qubits},
\doilink{10.1103/PhysRevA.92.022336}{Phys. Rev. A {\bf 92}, 022336 (2015)}.

\bibitem{Sedlacek2012}
J.A. Sedlacek,	A. Schwettmann,	H. K\"ubler,	R. L\"ow,	T. Pfau, and J.P. Shaffer,
\emph{Microwave electrometry with Rydberg atoms in a vapour cell using bright atomic resonances},
\doilink{10.1038/nphys2423}{Nat. Phys. {\bf 8}, 819 (2012)}.

\bibitem{Holloway2014}
C.L. Holloway, J.A. Gordon, A. Schwarzkopf, D. Anderson, S. Miller, N. Thaicharoen and G. Raithel,
\emph{Sub-wavelength imaging and field mapping via EIT and Autler-Townes splitting in Rydberg atoms},
\doilink{10.1063/1.4883635}{Appl. Phys. Lett. {\bf 104}, 244102 (2014)}.

\bibitem{Barredo2015} 
D. Barredo, H. Labuhn, S. Ravets, T. Lahaye, A. Browaeys, and C.S. Adams, 
\emph{Coherent Excitation Transfer in a ``Spin Chain'' of Three Rydberg Atoms}, 
\doilink{10.1103/PhysRevLett.114.113002}{Phys. Rev. Lett. {\bf 114}, 113002 (2015)}.

\bibitem{Schwarzkopf2013}
A. Schwarzkopf, D.A. Anderson, N. Thaicharoen, and G. Raithel
\emph{Spatial correlations between Rydberg atoms in an optical dipole trap},
\doilink{10.1103/PhysRevA.88.061406}{Phys. Rev. A {\bf 88}, 061406(R) (2013)}.

\bibitem{Barredo2014}
D. Barredo, S. Ravets, H. Labuhn, L. B\'eguin, A. Vernier, F. Nogrette, T. Lahaye and A. Browaeys,
\emph{Demonstration of a strong Rydberg blockade in three-atom systems with anisotropic interactions},
\doilink{10.1103/PhysRevLett.112.183002}{Phys. Rev. Lett. {\bf 112}, 183002 (2014)}.

\bibitem{Urban2009} 
E. Urban, T. A. Johnson, T. Henage, L. Isenhower, D. D. Yavuz, T. G. Walker, and M. Saffman, 
\emph{Observation of Rydberg blockade between two atoms},  
\doilink{doi:10.1038/nphys1178}{Nat. Phys. {\bf 5}, 110 (2009)}.

\bibitem{Gaetan2009} 
A. Ga\"{e}tan, Y. Miroshnychenko, T. Wilk, A. Chotia, M. Viteau, D. Comparat, P. Pillet, A. Browaeys, and P. Grangier 
\emph{Observation of collective excitation of two individual atoms in the Rydberg blockade regime}, 
\doilink{doi:10.1038/nphys1178}{Nat. Phys. {\bf 5}, 115 (2009)}.

\bibitem{Wilk2010} 
T. Wilk, A. Ga\"etan, C. Evellin, J. Wolters, Y. Miroshnychenko, P. Grangier, and A. Browaeys,  
\emph{Entanglement of Two Individual Neutral Atoms Using Rydberg Blockade},  
\doilink{10.1103/PhysRevLett.104.010502}{Phys. Rev. Lett. {\bf 104}, 010502 (2010)}.

\bibitem{Sackett2000}
C.A. Sackett, D. Kielpinski, B.E. King, C. Langer, V. Meyer, C.J. Myatt, M. Rowe, Q.A. Turchette, W.M. Itano, D.J. Wineland and C. Monroe,
\emph{Experimental entanglement of four particles},
\doilink{10.1038/35005011}{Nature {\bf 404}, 256(2000)}.

\bibitem{Gaetan2010}
A. Ga\"etan, C. Evellin, J. Wolters, P. Grangier, T. Wilk and A. Browaeys, 
\emph{Analysis of the entanglement between two individual atoms using global Raman rotations},
\doilink{doi:10.1088/1367-2630/12/6/065040}{New J. Phys. {\bf 12}, 065040 (2010)}.

\bibitem{Isenhower2010} 
L. Isenhower, E. Urban, X.L. Zhang, A.T. Gill, T. Henage, T.A. Johnson, T.G. Walker, and M. Saffman,
\emph{Demonstration of a Neutral Atom Controlled-NOT Quantum Gate},  
\doilink{10.1103/PhysRevLett.104.010503}{Phys. Rev. Lett. {\bf 104}, 010503 (2010)}.

\bibitem{Zhang2010}
X.L. Zhang, L. Isenhower, A.T. Gill, T.G. Walker, and M. Saffman,
\emph{Deterministic entanglement of two neutral atoms via Rydberg blockade},
\doilink{10.1103/PhysRevA.82.030306}{Phys. Rev. A {\bf 82}, 030306(R) (2010)}.

\bibitem{Jau2015}
Y.-Y. Jau,	A. M. Hankin,	T. Keating,	I. H. Deutsch, and	G.W. Biedermann,
\emph{Entangling atomic spins with a Rydberg-dressed spin-flip blockade},
\doilink{doi:10.1038/nphys3487}{Nat. Phys. {\bf 12}, 71 (2015)}.

\bibitem{Bouchoule2002}
I. Bouchoule and K. Moelmer,
\emph{Entangling atomic spins with a Rydberg-dressed spin-flip blockade},
\doilink{10.1103/PhysRevA.65.041803}{Phys. Rev. A {\bf 65}, 041803(R) (2002)}.

\bibitem{Pupillo2010}
G. Pupillo, A. Micheli, M. Boninsegni, I. Lesanovsky, and P. Zoller,
\emph{Strongly Correlated Gases of Rydberg-Dressed Atoms: Quantum and Classical Dynamics},
\doilink{10.1103/PhysRevLett.104.223002}{Phys. Rev. Lett. {\bf 104}, 223002 (2010)}.

\bibitem{Johnson2010}
J.E. Johnson and S.L. Rolston,
\emph{Interactions between Rydberg-dressed atoms},
\doilink{10.1103/PhysRevA.82.033412}{Phys. Rev. A {\bf 82}, 033412 (2010)}.

\bibitem{Keating2015}
Tyler Keating, Robert L. Cook, Aaron M. Hankin, Yuan-Yu Jau, Grant W. Biedermann, and Ivan H. Deutsch,
\emph{Robust quantum logic in neutral atoms via adiabatic Rydberg dressing},
\doilink{10.1103/PhysRevA.91.012337}{Phys. Rev. A {\bf 91}, 012337 (2015)}.

\bibitem{Zhang2012}
X.L. Zhang, A.T. Gill, L. Isenhower, T.G. Walker, and M. Saffman,
\emph{Fidelity of a Rydberg-blockade quantum gate from simulated quantum process tomography},
\doilink{10.1103/PhysRevA.85.042310}{Phys. Rev. A {\bf 85}, 042310 (2012)}.

\bibitem{Beguin2013} 
L. B\'eguin, A. Vernier, R. Chicireanu, T. Lahaye, and A. Browaeys,  
\emph{Direct Measurement of the van der Waals Interaction between Two Rydberg Atoms},
\doilink{10.1103/PhysRevLett.110.263201}{Phys. Rev. Lett. {\bf 110}, 263201 (2013)}.

\bibitem{Ravets2014} 
S. Ravets, H. Labuhn, D. Barredo, L. B\'eguin, T. Lahaye, and A. Browaeys, 
\emph{Coherent dipole-dipole coupling between two single atoms at an electrically-tuned F\"orster resonance}, 
\doilink{10.1038/NPHYS3119}{Nature Phys. {\bf 9}, (2014)}.

\bibitem{Gunter2012}
G. G\"unter, M. Robert-de-Saint-Vincent, H. Schempp, C. S. Hofmann, S. Whitlock, and M. Weidem\"uller,
\emph{Interaction Enhanced Imaging of Individual Rydberg Atoms in Dense Gases},
\doilink{10.1103/PhysRevLett.108.013002}{Phys. Rev. Lett. {\bf 108}, 013002}.

\bibitem{Zeiher2015} 
J. Zeiher, P. Schau{\ss}, S. Hild, T. Macr\`i, I. Bloch, and C. Gross, 
\emph{Microscopic Characterization of Scalable Coherent Rydberg Superatoms}, 
\doilink{10.1103/PhysRevX.5.031015}{Phys. Rev. X {\bf 5}, 031015 (2015)}.

\bibitem{Schauss2015} 
P. Schau{\ss}, J. Zeiher, T. Fukuhara, S. Hild, M. Cheneau, T. Macr\`i, T. Pohl, I. Bloch, and C. Gross, 
\emph{Crystallization in Ising quantum magnets}, 
\doilink{10.1126/science.1258351}{Science  {\bf 347}, 1455 (2015)}.

\end{thebibliography}
\end{document}